\documentclass[submission, Phys]{SciPost}
\usepackage{placeins}
\usepackage{lipsum}
\usepackage{xcolor}
\usepackage[normalem]{ulem}
\usepackage{comment}
\usepackage{tabularx}
\usepackage{graphicx}
\usepackage{dcolumn}
\usepackage{bm}
\usepackage{subcaption}
\usepackage{float}
\usepackage{booktabs}
\usepackage{dsfont}
\usepackage[ruled,vlined, linesnumbered]{algorithm2e}

\usepackage{bbm}
\usepackage{blindtext}
\usepackage{graphics}
\usepackage{verbatim}   
\usepackage{amsfonts}
\usepackage{amsmath}
\usepackage{amssymb}
\usepackage{adjustbox}
\usepackage{microtype} 
\allowdisplaybreaks 
\usepackage{xspace} 
\usepackage{xparse} 
\usepackage{multirow} 
\usepackage{tabstackengine}
\setstackEOL{\cr}

\usepackage{multirow}
\usepackage{physics2}
\usepackage{hyperref}
\hypersetup{colorlinks=true,breaklinks,linkcolor=blue,urlcolor=blue,citecolor=blue}
\usepackage{orcidlink}

\definecolor{darkgreen}{rgb}{0,0.5,0}
\definecolor{purple}{rgb}{0.6,0,0.5}
\definecolor{orange}{rgb}{1,0.5,0}
\definecolor{darkred}{rgb}{.7,0,0}
\definecolor{darkblue}{rgb}{0,0,.6}
\definecolor{grey}{rgb}{.6,.6,.6}
\definecolor{dimgreen}{rgb}{0.2,0.7,0.2}

\newcommand{\Eq}[1]{Eq.~\eqref{#1}}

\newcommand{\Fig}[1]{Fig.~\ref{#1}}

\newcommand{\mr}[1]{\mathrm{#1}}

\newcommand*{\ndots}{\kern-0.075em.\kern-0.05em.\kern-0.05em.}  
\newcommand*{\nidots}{.\kern-0.05em.\kern-0.05em.} 
\newcommand*{\ncdots}{\kern-0.15em\cdot\kern-0.2em\cdot\kern-0.2em\cdot\kern-0.15em}   

\NewDocumentCommand{\doubleI}{O{}}{\mathbbm{1}_{#1}}
\NewDocumentCommand{\doubleIb}{O{}}{{\overline{\mathbbm{1}}_{#1}}}
\NewDocumentCommand{\doubleIk}{O{}}{\mathbbm{1}^\ks_{\! #1}}
\NewDocumentCommand{\doubleId}{O{}}{\mathbbm{1}^\ds_{\! #1}}
\NewDocumentCommand{\doubleIp}{O{}}{\mathbbm{1}^\ps_{\! #1}}
\NewDocumentCommand{\doubleV}{O{}}{\mathbbm{V}_{\! #1}}
\NewDocumentCommand{\doubleVk}{O{}}{\mathbbm{V}^\ks_{\! #1}}
\NewDocumentCommand{\doubleVd}{O{}}{\mathbbm{V}^\ds_{\! #1}}
\NewDocumentCommand{\doubleVp}{O{}}{\mathbbm{V}^\ps_{\! #1}}
\NewDocumentCommand{\doublev}{o}{{\mathbbm{v}_{#1}}}
\NewDocumentCommand{\doubleVb}{o}{{\overline{\mathbbm{V}}_{\! #1}}}
\NewDocumentCommand{\doubleVt}{o}{{\widetilde{\mathbbm{V}}_{\! #1}}}
\NewDocumentCommand{\doubleVh}{o}{\widehat{{\mathbbm{V}}_{\! #1}}}
\NewDocumentCommand{\doubleW}{o}{\mathbbm{W}_{\! #1}}
\NewDocumentCommand{\doubleWk}{o}{\mathbbm{W}^\ks_{\! #1}}
\NewDocumentCommand{\doubleWd}{o}{\mathbbm{W}^\ds_{\! #1}}
\NewDocumentCommand{\doubleWb}{o}{{\overline{\mathbbm{W}}_{\! #1}}}
\NewDocumentCommand{\doubleWt}{o}{{\widetilde{\mathbbm{V}}_{\! #1}}}
\NewDocumentCommand{\doubleWh}{o}{{\widehat{\mathbbm{V}}_{\! #1}}}

\newcommand{\mL}{\mathcal{L}}

\newcommand{\mR}{\mathcal{R}}

\newcommand{\mN}{\mathcal{N}}
\newcommand{\bsigma}{{\boldsymbol{\sigma}}}

\newcommand{\chip}{\chi_{\text{p}}}
\newcommand{\ellb}{{\bar \ell}}
\newcommand{\Np}{N_{\text{p}}}
\newcommand{\Npar}{N_{\text{par}}}
\newcommand{\bk}{{\boldsymbol{k}}}

\newcommand{\bq}{\boldsymbol{q}}
\newcommand{\br}{\boldsymbol{r}}

\newcommand{\jvdomit}[1]{}

\begin{document}

\begin{center}
{\Large \textbf{
Adaptive Patching for Tensor Train Computations}}
\end{center}

\begin{center}

Gianluca Grosso\orcidlink{0009-0000-5445-3943}\textsuperscript{1,2*},
Marc K. Ritter\,\orcidlink{0000-0002-2960-5471}\textsuperscript{1,3},
Stefan Rohshap\,\orcidlink{0009-0007-2953-8831}\textsuperscript{4},
Samuel Badr\,\orcidlink{0009-0001-3090-1067}\textsuperscript{4},
Anna Kauch\,\orcidlink{0000-0002-7669-0090}\textsuperscript{4},
Markus Wallerberger\,\orcidlink{0000-0002-9992-1541}\textsuperscript{4},
Jan von Delft\,\orcidlink{0000-0002-8655-0999}\textsuperscript{1},
Hiroshi Shinaoka\,\orcidlink{0000-0002-7058-8765}\textsuperscript{5\textdagger}
\end{center}

\begin{center}
{\bf 1} Arnold Sommerfeld Center for Theoretical Physics, Center for NanoScience,
and Munich Center for Quantum Science and Technology,
Ludwig-Maximilians-Universit\"at M\"unchen, 80333 Munich, Germany
\\
{\bf 2} Institute of Physics, École Polytechnique Fédérale de Lausanne (EPFL), CH-1015 Lausanne, Switzerland
\\ 
{\bf 3} Center for Computational Quantum Physics, Flatiron Institute, 162~5th~Avenue, New~York, NY~10010, USA
\\
{\bf 4} Institute of Solid State Physics, TU Wien, 1040 Vienna, Austria
\\
{\bf 5} Department of Physics, Saitama University, Saitama 338-8570, Japan
\\
*\ gianluca.grosso@epfl.ch\\
\textdagger\ h.shinaoka@gmail.com
\end{center}

\begin{center}
\today
\end{center}

\section*{Abstract}
{\bf
Quantics Tensor Train (QTT) operations such as matrix product operator contractions are prohibitively expensive for large bond dimensions.
We propose an adaptive patching scheme that exploits block-sparse QTT structures to reduce costs through divide-and-conquer, adaptively partitioning tensors into smaller patches with reduced bond dimensions.
We demonstrate substantial improvements for sharply localized functions and show efficient computation of bubble diagrams and Bethe-Salpeter equations, opening the door to practical large-scale QTT-based computations previously beyond reach.
}

\vspace{10pt}
\noindent\rule{\textwidth}{1pt}
\tableofcontents\thispagestyle{fancy}
\noindent\rule{\textwidth}{1pt}
\vspace{10pt}

\section{Introduction}
\label{sec:intro}

In quantum many-body (QMB) physics, the central computational problem is the exponential growth of the Hilbert space with the number of particles, which corresponds to an exponential increase of required computational resources. This is an instance of the \emph{curse of dimensionality} whereby the computational resources required for the numerical solution of a problem often scale exponentially with its dimensionality.
In the QMB physics community, tensor networks (TNs), and in particular \textit{matrix product states} (MPSs), are well-established and standard techniques used to mitigate such an exponential increase of computational resources \cite{Fannes1992,White1992,Schollwock2011,Vidal2003,VerstraeteCirac2004,vonDelftTNNotes}. Such TN approaches are applicable whenever nominally exponentially costly functions are \textit{compressible}, in the sense that they can be represented by tensors of low rank.
There are several comprehensive MPS/TN software libraries~\cite{ITensors.jl, QSpace,tensorkitjl} in widespread use.

In parallel, tensor network approaches to other curse-of-dimensionality problems have been developed in the applied mathematics community \cite{Oseledets2009Intro, Kolda2009}. There, MPSs are known as \emph{tensor trains} (TTs), and thus in this work we will use ``tensor train'' and ``matrix product state'' interchangeably \cite{Oseledets2011}. Some TT algorithms developed in the applied mathematics community have recently been applied to problems in physics.
An example is the Tensor Cross Interpolation (TCI) algorithm~\cite{Oseledets2010,Dolgov2020}, which uses a low-rank TT for function approximation based on sampling the function in an active learning scheme.
It has been used, for instance, to compute the high-dimensional integrals that arise when evaluating Feynman diagrams, and was shown to converge more quickly than Monte Carlo methods for this purpose, while being more robust to the sign problem \cite{Fernandez2022}.
Quantics Tensor Trains (QTT) are another powerful tool, used for representing functions with scale separation
\cite{Oseledets2009, Khoromskij2011, Qttbook}, obtained by using binary representations for function variables and associating each bit with a tensor index.
This approach has found diverse applications across multiple fields of physics, including turbulent flow simulations~\cite{Gourianov2022-fg}, quantum field theoretical approaches~\cite{Hiroshi2023,Murray2024,Sroda2024-jm, Rohshap2025a, Rohshap2025b, Rohshap2025c}, high-resolution integration combined with TCI~\cite{Ritter2024}, quantum chemistry calculations~\cite{Jolly2024}, Gross--Pitaevskii dynamics~\cite{chen2025solvinggrosspitaevskiiequationquantic,boucomas2025quanticstensortrainsolving}, and correlated super-moir\'e systems~\cite{sun2025selfconsistenttensornetworkmethod}.

Despite these advances, most applications remain limited to proof-of-concept studies.
A primary reason is that the computational cost of these methods can be prohibitively high for large-scale applications.
For example, many applications require computing element-wise multiplication or convolution of functions, such as evaluating nonlinear terms in the Navier-Stokes equation~\cite{Gourianov2022-fg} or when solving many-body field-theoretical equations~\cite{Hiroshi2023,Rohshap2025a,Murray2024,Sroda2024-jm}.
Such operations require contracting two matrix product operators (MPO) of bond dimension $\chi$, resulting in $\mathcal{O}(\chi^4\mL)$ computational complexity that becomes prohibitively expensive for large $\chi$ ($\mL$ denotes the length of the MPO).
Furthermore, efficient parallelization of such contractions is non-trivial, leaving many modern, high-dimensional applications beyond the reach of standard numerical methods.
Consequently, there is a pressing need for efficient methods for function manipulation in QTT to enable practical large-scale applications.

A natural inspiration comes from adaptive mesh refinement (AMR)~\cite{Barth2005-jf}, a well-established and successful technique in computational fluid dynamics and related fields that adapts the mesh size to local solution features, improving accuracy while reducing computational cost.
A conceptually related strategy in numerical linear algebra is the $\mathcal{H}$-matrix or mosaic-skeleton approximation for matrices~\cite{Tyrtyshnikov1996, Gavrilyuk2002}, where a matrix is subdivided into smaller blocks and each block is approximated by a low-rank matrix.
The $\mathcal{H}$-matrix framework has been directly applied to quantum field theoretical calculations using nonequilibrium Green's functions~\cite{Kaye2021-ly}, where the time domain is partitioned into subintervals and treated within an $\mathcal{H}$-matrix representation.
Extending this matrix-based divide-and-conquer philosophy to the QTT setting, related non-adaptive partitioning strategies combined with QTT have already been applied to nonequilibrium Green's functions~\cite{inayoshi2025causalitybaseddivideandconqueralgorithmnonequilibrium}.

Motivated by these successes, we propose a similar \emph{divide-and-conquer} approach for function manipulation and interpolation in QTT.
Employing the quantics representation proposed in Ref.~\cite{Hiroshi2023}, our strategy is based on the concept of ``adaptive patching''. 
Our algorithm exploits the block-sparse structure of tensor cores that arises in QTT representations of functions with localized features.
It adaptively partitions the target tensor into smaller subtensors (``patches'') in a manner consistent with the binary encoding, representing each patch using a smaller bond dimension.
This allows us to perform TCI/SVD-based TT unfolding and other operations, such as MPO contractions, in a patch-wise manner.
In contrast to Ref.~\cite{Hiroshi2023}, the present work focuses on fully adaptive partitioning, in which the patch structure can in principle be refined dynamically and the patching order chosen more flexibly within the same framework. 

The remainder of this paper is organized as follows.
Section~\ref{sec:tt} provides a brief introduction to QTTs and establishes the necessary notation.
Section~\ref{sec:patching} presents our practical algorithm for adaptive patching and demonstrates its efficiency through numerical examples.
Section~\ref{sec:patched-mpo-mpo-contractions} introduces a divide-and-conquer approach for MPO--MPO contractions of patched QTT objects.
Sections~\ref{sec:bubble} and \ref{sec:BSE} present practical applications of our method.
Specifically, we demonstrate the computation of bubble diagrams in both imaginary-time and real-time formalisms in Sec.~\ref{sec:bubble}, and the solution of the Bethe-Salpeter equation in Sec.~\ref{sec:BSE}. Conclusions and an outlook are presented in Sec.~\ref{sec:conclusion}.

\section{\label{sec:tt}Tensor Train Formalism}
We briefly introduce the TT formalism and establish the necessary notation as a preparation for the subsequent sections, following Ref.~\cite{Fernandez2024}. A more detailed overview can be found there.

\subsection{Quantics Tensor Trains (QTT)}
A Quantics Tensor Train (QTT) is a function representation in terms of a tensor train whose indices encode the function argument in binary form.
To illustrate the idea, we consider a function $f(x)$ of a continuous variable $x \in [x_\text{min}, x_\text{max}]$, which we discretize on a uniform grid with $M = 2^\mR$ points.
The \(m\)th grid point is at position $x_m = x_\text{min} + m \Delta x$ with $m \in \{0, \ldots, M-1\}$ and $\Delta x = (x_\text{max} - x_\text{min}) / M$.
Then, in its quantics representation, the variable (linear index) $m$ is expressed in terms of the quantics indices $\sigma_{\ell}$, such that
\begin{equation}
    m = (\sigma_1 \sigma_2 \dots \sigma_\mR)_2 = \sum_{\ell=1}^{\mR} 2^{\mR-\ell} \sigma_{\ell}, \qquad \sigma_\ell \in \{0, 1\}.
\end{equation}
Here, each bit $\sigma_\ell$ represents an exponentially different length scale: the first bit $\sigma_1$ corresponds to the coarsest scale, dividing the domain into halves, while the last bit $\sigma_\mR$ corresponds to the finest scale.

We discretize $f(x)$ on this grid, yielding a vector $f_m=f(x_m)$ of length $M=2^\mR$.
Then, we reshape this vector into a tensor $F_{\sigma_1, \ldots, \sigma_\mR}$ of shape $2 \times 2 \times ... \times 2$ ($\mR$ times). This tensor is decomposed into a tensor train, also known as matrix product state (MPS), of the form 
\begin{equation}
    F_{\sigma_1 \ldots \sigma_\mR} \approx \widetilde{F}_{\sigma_1 \ldots \sigma_\mR} =
    \!\sum_{a_1=1}^{\chi_1}\!\ldots\!\sum_{a_{\mR-1}=1}^{\chi_{\mR-1}}
    [M_1]^{\sigma_1}_{1 a_1} [M_2]^{\sigma_2}_{a_1a_2} \cdots [M_\mR]^{\sigma_\mR}_{a_{\mR-1} 1},
    \label{eq:TTdecomposition}
\end{equation}
which can be graphically represented as 
\begin{equation}
    \raisebox{-5mm}{\includegraphics[scale=0.95]{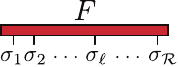}} \approx \raisebox{-5mm}{\includegraphics[scale=0.95]{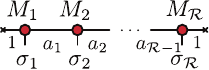}}.
    \label{eq:TTdecompositiondiagram}
\end{equation}
in tensor network notation.
Each tensor  \(M_\ell\) is of dimension $\chi_{\ell-1} \times 2 \times \chi_\ell$, where \(\sigma_\ell\) is the quantics bit index and $a_{\ell-1}$, $a_\ell$ are the virtual indices, or ``bonds'', summed over.
The bond dimensions $\chi_\ell$ quantify the amount of entanglement between the different length scales, and, thus, the information exchanged between those scales~\cite{Rohshap2025b, Rohshap2025c}. 

We can generalize this approach to a function of $\mN$ variables $f(\boldsymbol{x})$ by discretizing each variable on a binary grid with \(M=2^{\mR}\) points per variable.
This yields a tensor $F_\bsigma$ of shape $2 \times 2 \times ... \times 2$ ($\mL = \mN \times \mR$ times) with $\bsigma = (\sigma_{11}, \ldots, \sigma_{\mN 1}, \sigma_{12}, \ldots, \sigma_{\mN 2}, \ldots, \sigma_{1\mR}, \ldots, \sigma_{\mN \mR})$. Here, $\mR$ denotes the number of quantics bits per variable, so that the total number of QTT sites is $\mL = \mN\mR$.
Here, the index $\sigma_{n r}$ corresponds to the $r$-th bit of the $n$-th variable, and we order the indices by placing the bits of the same scale  side by side (\textit{interleaved representation}).
The resulting tensor of order $\mL$ is then factorized into a tensor train of length $\mL$ and local dimension 2 as in Eq.~\eqref{eq:TTdecomposition}.
Alternatively, we can group the bits of the same scale into a single external tensor leg, yielding a tensor train of length $\mR$ with local dimension $d=2^\mN$ (\textit{fused representation}). More generally, other orderings of the tensor indices are possible, such as placing all bits of each variable consecutively (\textit{normal ordering}), or applying the interleaved ordering only to a subset of variables while keeping the others in normal ordering; the most suitable ordering depends on the structure of the function to be approximated.



The amount of memory required for a QTT scales as \(\mathcal{O}(\chi_{\max}^2 \mL)\), with \(\chi_{\max} = \max_\ell \chi_\ell\). It scales most strongly with the bond dimensions \(\chi_\ell\) necessary to approximate the function \(f\) to the required accuracy, which is determined by the function's structure. 
In many practical applications, the functions of interest have been shown to be compressible as QTTs with small \(\chi_{\max}\), which does not increase with \(\mL\).
In that case, the computational cost of a QTT scales linearly in \(\mL\).
Upper bounds on \(\chi_{\max}\) for some classes of functions have been derived in Ref.~\cite{Lindsey2024}.

\subsection{Operations on Quantics Tensor Trains}
\begin{figure}[htbp]
    \centering
	\includegraphics{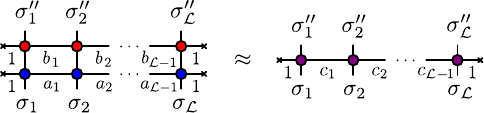}
	\caption{
    MPO--MPO contraction.
        Two MPOs are contracted by contracting their external legs, and the resulting MPO is compressed on the fly. Here, $\mL$ is the length of the MPOs.
    }
	\label{fig:MPOMPOcontr}
\end{figure}

In addition to function compression, operations such as convolution and Fourier transformation can be performed directly within the compressed QTT format, also at linear cost in $\mL$.
Such operations are implemented with matrix product operators (MPOs), also known as tensor train operators, which represent linear maps \(A_{\bsigma\bsigma'}\) acting on a QTT.
Instead of one external or physical leg per site-tensor of the QTT, MPOs have two external legs attached to each tensor, corresponding to \(\sigma_\ell\) and \(\sigma_\ell'\) at each site:
\begin{equation}
    \raisebox{-5.5mm}{\includegraphics{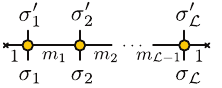}}.\label{eq:MPO}
\end{equation}
Diagonal MPOs represent element-wise (Hadamard) multiplication, while generic MPOs cover, e.g., convolutions and time-evolution operators.
These operations can be implemented by applying an MPO of bond dimension \(\chi_{\mathrm{op}}\) to a QTT of bond dimension \(\chi_{\max}\), or, when composing two operators, by contracting two MPOs (see Fig.~\ref{fig:MPOMPOcontr}).
Using standard ``zip-up''/fit-type contraction and on-the-fly recompression algorithms, the computational cost scales as \(\mathcal{O}(\chi_{\max}^3\,\chi_{\mathrm{op}}\,\mL)\): it reduces to \(\mathcal{O}(\chi_{\max}^4\,\mL)\) in the generic case \(\chi_{\mathrm{op}}\sim \chi_{\max}\), but becomes cubic when \(\chi_{\mathrm{op}}\) does not scale with \(\chi_{\max}\)~\cite{Schollwock2011}.
For quantics ``superfast'' Fourier transformation, the QFT MPO has a small, constant bond dimension (in our examples, \(\chi_{\mathrm{op}}\approx 15\) already reaches machine precision), resulting in cubic scaling~\cite{Dolgov2012}. 
A detailed description of various operations on QTTs can be found in Sec.~6.2 of Ref.~\cite{Fernandez2024}.

\subsection{Tensor Cross Interpolation}
Tensor Cross Interpolation (TCI) is an algorithm that constructs a TT decomposition of a tensor \(F\) by adaptively sampling a subset of tensor elements~\cite{Oseledets2010, Dolgov2020, Fernandez2022, Fernandez2024}.
TCI is initialized with a TT of very small rank, typically \(\chi_\ell = 1\). If there is prior knowledge about \(F\), points that are known to be important can be included in the initial TT.
These points are called \emph{initial pivots}.
The algorithm then optimizes the TT iteratively, by identifying points where the current TT approximation deviates significantly from \(F\), and updating the TT by introducing these points as new pivots.
For each pivot \(\hat\bsigma\), all points that differ from \(\hat\bsigma\) in at most one digit are reproduced exactly. The points that differ from \(\hat\bsigma\) in two digits are sampled for new pivots to be introduced in the next iteration, using one of several possible sampling strategies \cite{Fernandez2024}.
The algorithm terminates when the estimated TT approximation error \(\epsilon\) falls below a prescribed tolerance $\tau$.
The (normalized) error estimate is defined as
$\epsilon \equiv \frac{\|\widetilde{F}_{\bsigma} - F_{\bsigma}\|_\infty}{\|F_{\bsigma}\|_\infty}
= \frac{\max_{\bsigma} |\widetilde{F}_{\bsigma} - F_{\bsigma}|}{\max_{\bsigma} |F_{\bsigma}|}$ over all sampled points $\bsigma$.
Unlike SVD-based methods, TCI does not require sampling the entire tensor, reading only $\mathcal{O}(\chi_{\max}^2 \mL)$ tensor elements, which can make it exponentially faster than SVD-based TT decomposition.
However, this sampling procedure implies that TCI cannot guarantee a global error bound on the approximation error across all \(\bsigma\), as this is impossible without sampling all points and thus returning to the exponential scaling the algorithm seeks to avoid.
For more details on the TCI algorithm, we refer to Refs.~\cite{Fernandez2022, Fernandez2024}.

\section{\label{sec:patching}Adaptive patching}
In this section, we briefly introduce the adaptive patching concept proposed in Ref.~\cite{Hiroshi2023}, and describe a practical and heuristic algorithm. 
Then, we illustrate its behavior on representative examples of tensors with localized structures. 
Although we focus on the QTT representation, the patching scheme is applicable to any TT representation.

\subsection{Motivation: block-sparse structures}
To motivate the patching concept, consider a sum of $N_\mathrm{G}$ delta functions in a high-dimensional configuration space:
\begin{equation}
f(\bsigma) = \sum_{g=1}^{N_\mathrm{G}} \delta_{\bsigma, \bar{\bsigma}^g},
\end{equation}
where $\bar{\bsigma}^g = (\bar{\sigma}^g_1, \ldots, \bar{\sigma}^g_\mL)$ specifies the position of the $g$-th peak in the quantics representation.

Each individual delta function can be exactly represented by a TT with bond dimension 1.
When the $N_\mathrm{G}$ peaks are located at random, spatially separated positions in the high-dimensional configuration space, there is no correlation or compressibility between different peaks.
Therefore, the TT representation of the superposition requires bond dimension $\chi = N_\mathrm{G}$, and the resulting TT exhibits a block-sparse structure.
Specifically, for a sum of delta functions $f(\bsigma) = \sum_{g=1}^{N_\mathrm{G}} \delta_{\bsigma, \bar{\bsigma}^g}$, the site tensors take the form:
\begin{align}
[M_\ell]_{\sigma_\ell}^{m_{\ell-1}m_\ell} &= \sum_{g=1}^{N_\mathrm{G}} \delta_{m_{\ell-1}, g} \delta_{\sigma_\ell, \bar{\sigma}^g_\ell} \delta_{m_\ell, g},
\end{align}
where the bond dimension at each site is $\chi_\ell = N_\mathrm{G}$.
This structure is highly sparse: most elements of the tensor $[M_\ell]_{\sigma_\ell}^{m_{\ell-1}m_\ell}$ are zero, with only $\mathcal{O}(N_\mathrm{G})$ nonzero entries out of $\mathcal{O}(\chi_{\max}^2) = \mathcal{O}(N_\mathrm{G}^2)$ total elements.

While this example of delta functions represents an extreme case, more realistic functions with localized features (such as narrow Gaussians) exhibit qualitatively similar block-sparse structures, where many blocks of the TT tensors are nearly zero.
The adaptive patching scheme is designed to exploit this sparsity by subdividing the configuration space so that each patch captures a single (or few) localized feature(s) with a smaller bond dimension $\chip \ll N_\mathrm{G}$.
This results in reduced memory cost with a scaling of only $\mathcal{O}(N_\mathrm{G})$ instead of $\mathcal{O}(N_\mathrm{G}^2)$.

\subsection{Patching a tensor} \label{sec:patching-tensor}
We first introduce the notion of a \emph{patch} of an $\mL$-dimensional tensor
\begin{equation}
 F_{\bsigma} = \raisebox{-5mm}{\includegraphics[scale=0.95]{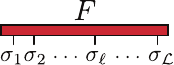}}.
\end{equation}
A patch ${F}^{p_\ell}_{\bsigma}$ is a projection of the original tensor onto a particular subdomain, defined by index \(\sigma_\ell\) matching \(p_\ell\), i.e. 
\begin{equation}
    F^{p_{\ell}}_{\bsigma} =
    F_{\bsigma} \delta_{\sigma_{\ell}}^{p_{\ell}} =
    \begin{cases}
        F_{\bsigma} & \text{if } \sigma_{\ell} = p_{\ell}, \\
        0 & \text{otherwise}.
    \end{cases}
    \label{eq:tensorSliceSingle}
\end{equation}
We can then perform a TT approximation of the patched tensor $F^{p_\ell}_{\bsigma}$, which can be cast into the form
\begin{equation}
    \widetilde{F}^{p_\ell}_{\bsigma} = \raisebox{-5mm}{\includegraphics[scale=0.95]{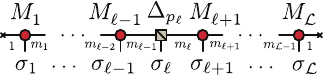}},
    \label{eq:projectedtensortrain}
\end{equation}
where we define the projection tensor \(\Delta_{p_\ell}\)
as
\begin{equation}
    \raisebox{-4.5mm}{\includegraphics[scale=0.95]{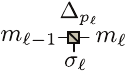}} =
    [\Delta_{p_\ell}]_{m_{\ell-1}m_\ell}^{\sigma_\ell} =
    \delta_{\sigma_\ell p_\ell} \delta_{m_{\ell-1}m_\ell} .
\end{equation}
This procedure can be generalized to projections on multiple indices \(p_{\ell_1} \ldots p_{\ell_N}\), where \(\ell_1 \ldots \ell_N\) have arbitrary, but unique (\(\ell_i \neq \ell_j\) for \(i \neq j\)), values in arbitrary order.
We define a tensor projected on multiple indices as
\begin{equation}
    F^{p_{\ell_1}\ldots p_{\ell_N}}_{\bsigma} =
    F_{\bsigma} \delta_{\sigma_{\ell_1}}^{p_{\ell_1}} \cdots \delta_{\sigma_{\ell_N}}^{p_{\ell_N}} =
    \begin{cases}
        F_{\bsigma} & \text{if } \sigma_{\ell_1} = p_{\ell_1}, \ldots, \sigma_{\ell_N} = p_{\ell_N},  \\
        0 & \text{otherwise}.
    \end{cases}
    \label{eq:tensorSliceMulti}
\end{equation}
With a complete set of disjoint subdomains \(F_{\bsigma}^{p_{\ell_1} \ldots p_{\ell_N}}\), the full tensor is recovered by summation over all patches \(F_{\bsigma} = \sum_{p_{\ell_1}} \cdots \sum_{p_{\ell_N}} F_{\bsigma}^{p_{\ell_1}\ldots p_{\ell_N}}\).

Analogously to \Eq{eq:projectedtensortrain}, a TT approximation can be constructed for each patch \(F_{\bsigma}^{p_{\ell_1} \ldots p_{\ell_N}}\), which contains one projection tensor \(\Delta\) per projected index.
An approximation of the full tensor can then be obtained by a sum over the TT approximations for all subdomains,
\begin{equation}
	F_{\bsigma} \simeq 
     \sum_{p_{\ell_1}= 1}^{d_{\ell_1}} \cdots \sum_{p_{\ell_N}= 1}^{d_{\ell_N}} \widetilde{F}_{\bsigma}^{p_{\ell_1} \dots p_{\ell_N}}.
	\label{eq:NslicesSum}
\end{equation}
As demonstrated in the examples below, the bond dimensions of a TT approximation for each patch may be much smaller than the bond dimension of a TT approximating the full tensor. The algorithms presented in this work aim to exploit this reduction in bond dimension for reduced memory consumption and runtime by setting up a suitable set of patches.

\subsection{Adaptive patching algorithm}\label{sec:adaptivePatchingAlgorithm}

The adaptive patching algorithm generates a set of patches according to the structure of the tensor to be approximated.
It can be combined with any scheme that generates TT approximations of tensors such as SVD or TCI.
In the following, we sketch the algorithm in a general way.
For its specific implementation in combination with Tensor Cross Interpolation on quantics grids, called \textit{patched Quantics Tensor Cross Interpolation} (pQTCI), see App.~\ref{app:pqtci}.

For simplicity, we patch a tensor from the first index to the last index.
The algorithm can be generalized to patch a tensor from any index to any other index.
Given a maximum rank per patch of \(\chip\), henceforth called ``bond-dimension cap'' or simply ``bond-cap'', and a tolerance \(\tau\), the algorithm proceeds as follows:
\begin{enumerate}
    \item Approximate the tensor with a TT \(\tilde{F}_\bsigma\) having maximum rank  \(\chip\).
    \item Check whether the estimated error \(\bigl\|F_\bsigma - \widetilde{F}_\bsigma\bigr\|\), in an appropriate norm, is smaller than the tolerance \(\tau\). If it is, the algorithm terminates and returns \(\widetilde{F}_\bsigma\).
    \item Otherwise, pick the next index \(\sigma_\ell\) in the predetermined order of indices to be projected.
    For each value of \(p_\ell = 1, \ldots, d_\ell\), call the algorithm \textit{recursively} on the patch \(F^{p_\ell}_\bsigma\).
    Collect the return values of each call, i.e., a collection of projected TTs, and return the resulting collection.
\end{enumerate}
\begin{figure}
	\centering \includegraphics{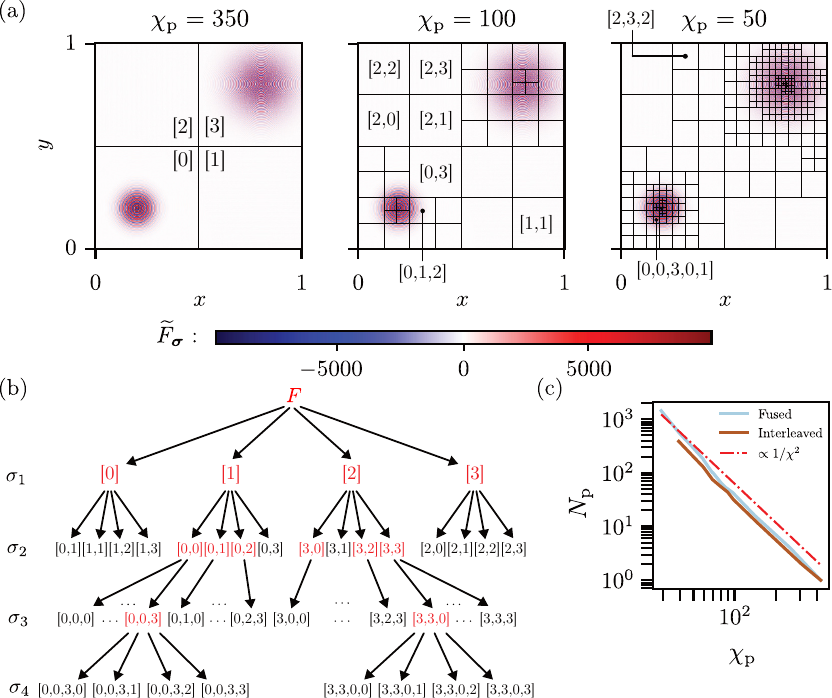}
	\caption{
    (a) 
    Partitioning of the domain of the bivariate function $f(\boldsymbol{r})$, Eq.~(\ref{eq:localFunc}), as obtained from the adaptive patching algorithm. We used the fused representation with local dimension $d\!=\!4$. 
    Smaller maximum bond dimensions per patch $\chip$ yield finer partitionings of the domain.
    (b) Hierarchical tree that records the patch refinement: each tensor slice is attached to its parent, and red leaves mark the yet-to-converge patches produced by the intermediate steps of the refinement.
    (c) Number of total patches vs.\ bond-dimension cap $\chip$. 
    The scaling $\Np \sim \chip^{-2}$ indicates that the number of total parameters is approximately constant for different choices of $\chip$, for both fused and interleaved representations (see the main text for more details).
    }
	\label{fig:patchEvolution+AlgTree}
\end{figure}

The algorithm creates a \textit{tree structure} of patches, each of which is a projected tensor train of the form
\begin{equation}
	\raisebox{-4.5mm}{\includegraphics{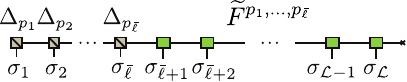}},
	\label{eq:patchedTT}
\end{equation}
where both the \textit{prefix} indices $[p_1,\dots,p_{\bar{\ell}}]$ and the length of the prefix -- the \textit{patching level} -- $\bar{\ell} \in \{1,\dots,\mL\}$ can be different for each patch.
By construction, the algorithm produces a collection of $\Np$ strictly non-overlapping patches that collectively cover the entire configuration space of the original tensor.

When refining a patch during the recursive application of the adaptive patching algorithm, we can truncate the patch TT to a certain bond dimension using either the maximum norm or the Frobenius norm.
The maximum norm is typically used in TCI, while the Frobenius norm is used in SVD.
In our numerical examples with TCI, we employ a global relative error criterion $\bigl\|F_\bsigma - \widetilde{F}_\bsigma\bigr\|_\infty < \tau \bigl\| F_\bsigma \bigr\|_\infty$, where the tolerance $\tau$ is applied uniformly across all patches.
This global normalization prevents the truncation from being dominated by noise in patches where the function takes small values.
A similar global measure can be used for Frobenius-norm truncation when performing adaptive patching with SVD (not shown in the present paper).

We illustrate the behavior of the adaptive patching algorithm through a simple, representative example, shown in Fig.~\ref{fig:patchEvolution+AlgTree}(a), which we also employ to examine the dependence of $\Np$ on the fixed parameter $\chip$. 
We consider the function describing two oscillating Gaussians,
\begin{gather}
		f(\boldsymbol{r}) =
        A \, e^{-\frac{(\boldsymbol{r} - \boldsymbol{r}_1)^2}{2\sigma^2}}
        \sin(k|\boldsymbol{r}- \boldsymbol{r}_1|)
        +
        \frac{A}{2}\, e^{-\frac{(\boldsymbol{r} - \boldsymbol{r}_2)^2}{\sigma^2/2}}
        \sin(k|\boldsymbol{r}- \boldsymbol{r}_2|/2),
		\label{eq:localFunc}
\end{gather}
where we set
\(
	\boldsymbol{r}_{1} = ( 0.2, 0.2),
    \boldsymbol{r}_{2} = ( 0.8, 0.8),
	A = 10^4,
    \sigma = 10^{-1},
\) and 
\(
    k = 10^3
\).
We discretize \(f\) on the domain \([0, 1]^2\) with \(\mR = 17\) fused quantics bits, and compress it with a tolerance of \(\tau = 10^{-7}\) using adaptive patching.
Varying the bond-cap \(\chip\) allows us to track how the domain is successively partitioned into smaller patches as \(\chip\) decreases.
For example, setting \(\chip = 350, 100, 50\) results in the patch structures shown in panels \Fig{fig:patchEvolution+AlgTree}(a).
The corresponding hierarchical refinement tree for the second case, \(\chip = 100\), is shown in panel (b).
The total patch count $\Np$ sensitively depends on the chosen bond-cap $\chip$. 
As shown in panel (c), the number of patches is approximately \(\Np \sim \chip^{-2}\) for the oscillating Gaussians considered here. Since memory consumption scales with \(\mathcal{O}(\chip^2 \Np)\), it stays approximately constant for different choices of \(\chip\). This is not generically the case, and some examples shown below have a different relation between \(\Np\) and \(\chip\).

\subsection{Computational complexity}
Table \ref{tab:patchingComplexity} summarizes the computational complexity of the adaptive patching scheme in terms of memory and runtime.
Assuming each patch is expressed as a TT of rank $\chip$,
the number of parameters \(\Npar\) is given by $\mathcal{O}(\Np \chip^2 d \mL)$, where $\Np$ is the number of patches defined earlier and $d$ is the local dimension.

The runtime cost depends on whether the TT is constructed by SVD or TCI and we will focus on the latter in the following. It
is characterized by two factors: the number of function evaluations and the construction time of the TT through linear algebra operations on the sampled function values.
Each construction of a TT of bond dimension $\chip$ requires $\mathcal{O}(\chip^2 d^2 \mL)$ function evaluations, and the number of such constructions needed scales with the number of patches.
Therefore, the total number of function evaluations is $\mathcal{O}(\Np \chip^2 d^2 \mL)$.
Each construction of a TT of bond dimension $\chip$ requires $\mathcal{O}(\chip^3 d^2 \mL)$ floating-point operations, resulting in a total runtime cost of $\mathcal{O}(\Np \chip^3 d^2 \mL)$.
When the function is expensive to evaluate, the runtime cost can be dominated by the number of function evaluations.

The above estimates for function evaluations and algebraic operations are based on a full pivoting strategy with two-site updates~\cite{Fernandez2024}.
These become $\mathcal{O}(\Np \chip^2 d \mL)$ and $\mathcal{O}(\Np \chip^3 d \mL)$, respectively, if a rook pivoting strategy, which is more efficient for large local dimension $d$, is used instead.
For more details, refer to Table 2 and related discussion in Ref.~\cite{Fernandez2024}.


\begin{table}
    \centering
    \begin{tabular}{ccc}
      \toprule
        \textbf{Number of parameters} & \textbf{Function evaluations} & \textbf{Algebra operations} \\
        \midrule
        $\mathcal{O}(\Np \chip^2 d \mL)$ &
        $\mathcal{O}(\Np \chip^2 d^2 \mL)$ &
        $\mathcal{O}(\Np \chip^3 d^2 \mL)$ \\
        \bottomrule
    \end{tabular}
    \caption{
        Computational complexity of the adaptive patching scheme.
        The function evaluations and algebraic operations are based on TCI with a full pivoting strategy with 2-site updates~\cite{Fernandez2024}.
    }
    \label{tab:patchingComplexity}
\end{table}

\subsection{Realistic example: 2D Green's functions}
We now present a concrete application of the adaptive patching algorithm, using its TCI-based variant, pQTCI, mentioned above. We consider the toy Green's function  

\begin{equation}
  G(\bk)
  \;=\;
  \frac{1}
       {\;\omega+\mu-\varepsilon_{\bk}+i\delta\,},
  \label{eq:2DGreen}
\end{equation}
where the non-interacting dispersion is taken as
\(\varepsilon_{\bk}=-2\cos k_{x}-2\cos k_{y}\)  
and we set the chemical potential to \(\mu=0\) and take $\omega=0.1$.
The function in~\eqref{eq:2DGreen} can either be understood as a retarded Green's function of the two-dimensional Hubbard model with broadening $\delta$ evaluated at a specified (real) frequency $\omega$, or as a Matsubara Green's function  $G(\bk,i\nu)$, evaluated at an effective chemical potential
of $\mu_\mr{eff} = 0.1$ and a specified Matsubara frequency $\nu = \delta$, which thereby encodes a temperature-dependent parameter.
The parameter \(\delta\) controls how sharply the Green's function is localized: as \(\delta\to 0\) the poles narrow and \(G(\bk)\) becomes increasingly singular.
Such sharp structures on a curved surface increase the QTT bond dimension at low temperatures, roughly as $\chi \propto \delta^{-1/2}$, as indicated by the numerical data in \cite{Ritter2024}.
This makes the function an ideal testbed for patched QTCI.

We discretize \(G(\bk)\) by quantics rebasing,  
$
  G(\bk)
  \longrightarrow
  G_{\bsigma} = G \bigl(\bk(\bsigma)\bigr),
$
with bit string
\(\boldsymbol{\sigma}=(\sigma_{1},\dots,\sigma_{\mR})\) in the fused
ordering, or \(\boldsymbol{\sigma}=(\sigma_{k_x1},\sigma_{k_y1},\dots,\sigma_{k_x\mR},\sigma_{k_y\mR})\) in the interleaved ordering.  
Throughout we use \(\mR=15\) bits per \(\bk\)-component.

After fixing a bond-cap \(\chip\) and a global tolerance \(\tau=10^{-7}\), we assess the convergence of the patched approximation by monitoring the pointwise discrepancy
\begin{equation}
  \text{err}(\bk)
  = \bigl|\operatorname{Re}\widetilde G_{\textrm{TCI}}(\bk)
                -\operatorname{Re}\widetilde G_{\textrm{pQTCI}}(\bk)\bigr|,
  \label{eq:localError2DGreen}
\end{equation}
defined over the entire Brillouin zone \( [-\pi,\pi]^2 \).
This error measure should be interpreted as a \emph{relative} indicator: it quantifies the deviation of the pQTCI approximation from a reference obtained via a standard TCI representation of the same function. As such, it validates the internal consistency of the comparison between the patched and unpatched approaches presented below, but it does not by itself guarantee absolute convergence of the underlying function approximation.

To evaluate the approximate Green's function on a uniform \(\bk\)-grid, we proceed as follows.
For each momentum point \(\bk\), we compute the corresponding multi--index \(\bsigma(\bk)\) under the QTT mapping.
The value of \(\operatorname{Re}G(\bk)\) is then obtained by evaluating the QTT associated with the unique patch \(\operatorname{Re}\bigl(\widetilde{G}^{p_1,\dots,p_{\ellb}}\bigr)\) such that $\bigl(\bsigma(\bk)\bigr)_1 = p_1 \wedge \cdots \wedge \bigl(\bsigma(\bk)\bigr)_\ellb = p_\ellb$.
This procedure is repeated independently for all grid points.

\begin{figure}[htbp]
    \centering 
    \includegraphics[width=\linewidth]{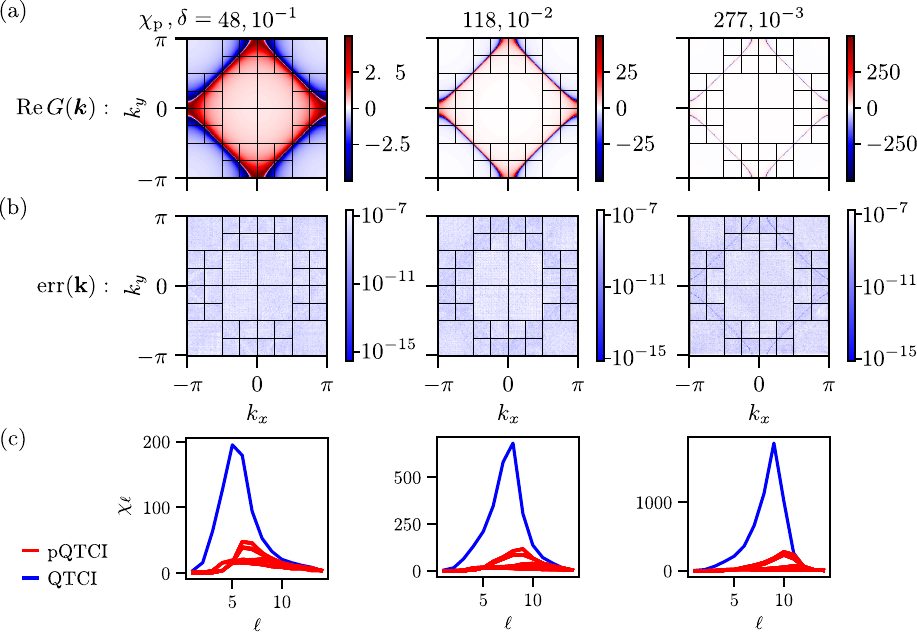}
    \caption{
        Patched QTCI approximation of
           \(\operatorname{Re}G(\bk)\) defined in Eq.~\eqref{eq:2DGreen}.
           (a) Heatmaps of the patched tensor train evaluated on \([-\pi,\pi]^{2}\) for bond-caps \(\chip=48,118,277\) (corresponding to
           \(\delta=10^{-1},10^{-2},10^{-3}\), respectively) at $\tau=10^{-7}$.
           (b) 
           Pointwise error $\text{err}(\bk)$ [Eq.~\eqref{eq:localError2DGreen}] for the same patched approximations, plotted using a $\log_{10}$ color scale.
           (c) Comparison of the bond-dimension profiles for the pQTCI (one line per patch) and standard QTCI representations of \(\operatorname{Re}G(\bk)\). Every patch in the pQTCI result adheres to the prescribed bond-cap $\chip$.
    }
    \label{fig:2DGreenErrorHeatmap}
\end{figure}

Figure~\ref{fig:2DGreenErrorHeatmap} illustrates the patched QTCI approximation of the real part of the Green's function for three values of the broadening parameter \(\delta\), analyzes its convergence behavior, and compares the results with those obtained from a standard QTCI routine.
The top panels, Fig.~\ref{fig:2DGreenErrorHeatmap}(a), show the reconstructed \(\operatorname{Re}G(\bk)\) obtained from the patched tensor--train representation, while the middle panels, Fig.~\ref{fig:2DGreenErrorHeatmap}(b), display the local error \(\text{err.}(\bk)\), as defined in~\eqref{eq:localError2DGreen}, over the entire Brillouin zone.
The fact that the error remains within the prescribed tolerance across the domain indicates that the patched approximation yields results consistent with those obtained from the standard TCI representation.
Finally, Fig.~\ref{fig:2DGreenErrorHeatmap}(c) reports the bond--dimension distributions for both the patched and single--TT approximations of \(\operatorname{Re}G(\bk)\) at the three bond-caps
$\chip = 48$, $118$, and $277$,
corresponding to $\delta = 10^{-1}$, $10^{-2}$, and $10^{-3}$, respectively.
The patched approach exhibits a pronounced rank reduction relative to the single QTCI tensor--train representation.
Moreover, two distinct classes of patches emerge: large, low--rank patches cover the  regions of the Brillouin zone where the Green's function is relatively smooth, while smaller, higher--rank patches are required to resolve the sharply structured, nontrivial features of the Green's function, as illustrated by the two groups of red lines in panels Fig.~\ref{fig:2DGreenErrorHeatmap}(c).

\begin{figure}[htbp]
    \centering
    \includegraphics{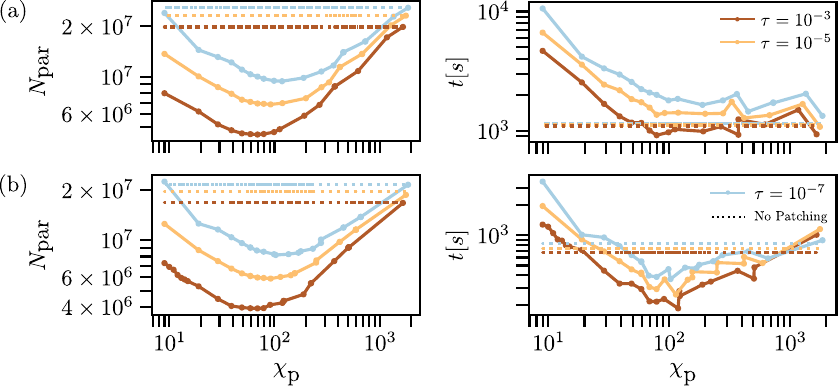}
    \caption{
        Total parameter count (left) and CPU run time (right) versus bond-cap \(\chip\) of a patched QTCI approximation of $\text{Re}\left(G(\bk)\right)$ for \(\delta=10^{-3}\), using interleaved (a) or fused (b) QTT index ordering. Dotted lines show the corresponding standard-QTCI measurements.
        }
    \label{fig:memoryTime2DGreen}
\end{figure}

We now benchmark patched QTCI (pQTCI) against the standard QTCI routine for the Green's function introduced above. Fixing the value of the broadening
\(\delta = 10^{-3}\) we measure the \emph{run time} on an Intel Xeon W-2245 CPU @ 3.90 GHz, and the \emph{memory footprint}, here defined as the total number of floating-point parameters in all patches \(\operatorname{Re}\widetilde{G}^{p_{1},\dots,p_{\bar\ell}}\). We scan the bond-cap \(\chip\) and compare fused and interleaved bit orderings at different set tolerances as illustrated in Fig.~\ref{fig:memoryTime2DGreen}. 

Each curve exhibits an optimal value $\chi_{\text{p}}^{\mathrm{best}}$, corresponding to the minimum of the respective curve, with generally distinct optima for the parameter-count and run-time metrics. Moreover, a modest but systematic performance advantage is observed for the fused representation, most prominently in terms of run time. This behavior can be attributed to the particular structure of the Green's function, whose symmetry properties are naturally aligned with the domain splitting induced by the fused representation. As a result, the adaptive patching procedure requires fewer refinement steps to capture the dominant features of the function, leading to a slight reduction in the overall computational effort.

Focusing on the parameter-count panels of Fig.~\ref{fig:memoryTime2DGreen}, one observes that the curves converge to the number of parameters obtained by the single TCI approximation in both the small- and large-$\chip$ regimes. For small values of $\chip$, this behavior is a consequence of \emph{overpatching}, as discussed in the subsequent section, whereby excessive domain subdivision offsets the reduced bond-dimension in each patch. In contrast, for sufficiently large $\chip$, the bond-cap on each patch reaches or exceeds the maximal bond dimension obtained for a standard TCI approximation of $\operatorname{Re}G(\bk)$. In this regime, the patched construction effectively collapses to a single patch, reproducing the parameter count of the unpatched TCI representation.

\begin{figure}[htbp]
    \centering
    \includegraphics{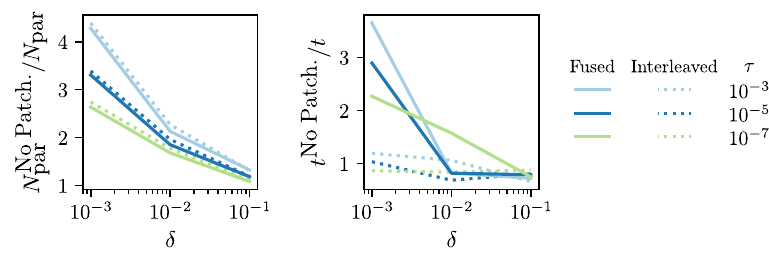}
    \vspace{-1em}
    \caption{Parameter and run-time ratios between the best patched approximation (at \(\chip^{\mathrm{best}}\)) and a single-TT QTCI
    approximation, as a function of the broadening \(\delta\).} 
    \label{fig:deltavsMemoryTime}
\end{figure}

A global perspective on the ``return on investment'' is provided in Figure~\ref{fig:deltavsMemoryTime}, which plots, for both parameter count and runtime, the ratio of the standard QTCI result to the best patched QTCI result at the optimal bond-cap $\chip^{\mathrm{best}}$ as a function of $\delta$.
A larger ratio indicates a greater benefit from using pQTCI.
For the broadest case, $\delta = 10^{-1}$, pQTCI offers no improvement: the Green's function is already smooth enough that a single TT is adequate, and additional patching only adds runtime overhead.
As $\delta$ decreases and the Green's function develops sharper features, adaptive domain partitioning concentrates computational effort where it is most needed, leading to increasing savings from pQTCI.

\subsection{Overpatching}
\label{sec:overpatching}
\begin{figure}[t]
	\centering 
\includegraphics{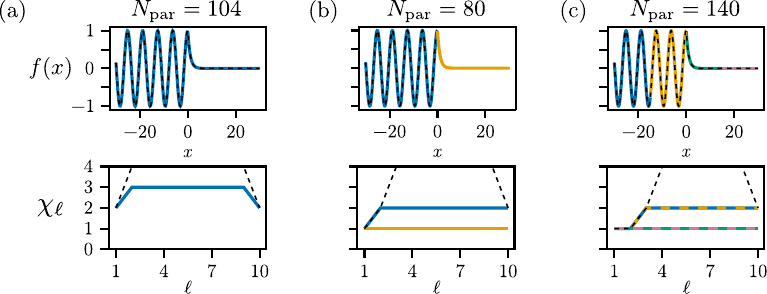}
\caption{
        Domain partitioning of the function $f(x) = \cos(x)\text{ if } x<0,\text{ else }\exp(-x)$ and QTT bond dimensions; colors link each segment (top panels) to the bond dimensions of its TT unfolding (bottom panels).
        Panels (a)--(c) show a single-domain approximation, an optimal subdivision, and an over-subdivided case; each panel is annotated with the total TT parameter count $\Npar$.}
	\label{fig:1DOverpatchingBonddim}
\end{figure}

The bond-cap $\chip$ plays a central role in the efficiency of the adaptive patching scheme: it must be chosen such that the resulting approximation is genuinely more economical than a direct (Q)TCI or SVD unfolding, e.g., in memory and/or in time for unfolding or subsequent MPO--MPO contractions (see Sec.~\ref{sec:patched-mpo-mpo-contractions}).
The maximum TT rank $\chi$ obtained from a standard run can serve as an initial guideline for selecting $\chip$. However, this information is often unavailable, and even when $\chi$ can be estimated---e.g., from a previous but computationally expensive TCI/SVD attempt---the optimal value of $\chip$ remains highly problem-dependent.
Figure~\ref{fig:1DOverpatchingBonddim} illustrates this point with a toy example. 

The target function consists of two pieces, an oscillatory segment for $x < 0$ followed by an exponential tail for $x > 0$. The colormap highlights that some domain partitions are clearly more efficient than others. For example, in the rightmost panels, Fig.~\ref{fig:1DOverpatchingBonddim}(c), where the domain is subdivided too aggressively, the number of resulting patches becomes so large that the overall cost surpasses that of a single, full-domain approximation. Indeed, splitting either the exponential tail or the oscillatory region in half produces multiple copies of the same tensor train, representing the same oscillatory or exponential features on smaller subdomains. These additional patches offer no computational advantage and merely increase the total parameter count.

This phenomenon---hereafter referred to as \textit{overpatching}---commonly arises either when the patching scheme is applied to functions with no sharply localized structure or when the prescribed bond-cap $\chip$ is chosen too small for the problem at hand. In both cases, once a patch reaches the bond-dimension limit, the algorithm’s adaptivity mechanism triggers further domain refinement, regardless of whether significant variation is present within that patch. The result is a proliferation of sub-patches with reduced spatial extent but tensor-train representations that inherit essentially the same bond dimensions as their parent patch, thereby undermining the intended efficiency gains of the subdivision.

\begin{figure}[htbp]
	\centering
	\includegraphics{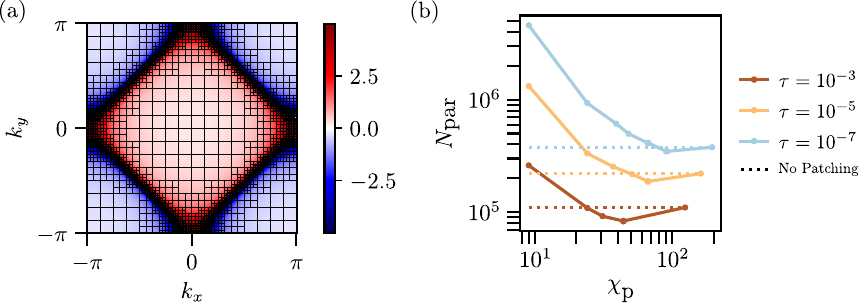}
	\caption{Overpatching for 2D Green's function. (a) Example of the domain subdivision produced with $\mR = 15$, $\chip = 9$, $\tau = 10^{-7}$, $\delta=10^{-1}$ and $\omega = 10^{-1}$. The domain partition is highly redundant. (b) Total number of TT parameters returned by pQTCI (solid curves) at three target tolerances, plotted against the bond-cap $\chip$.  For comparison, the parameter count of a full-domain TCI compression is shown as a dotted line. For excessively small bond-caps $\chip$, overpatching inflates the parameter count beyond the number of parameters required for a single TT.
    }
	\label{fig:2DGreenOverpatching}
\end{figure}

Figure~\ref{fig:2DGreenOverpatching} illustrates a failure mode of patched QTCI when applied to the two-dimensional Green's function of Eq.~\eqref{eq:2DGreen} with fixed \(\delta = 10^{-1}\) (another example with a different function is shown in App.~\ref{app:overpatching}). 
If the patching threshold \(\chip\) is chosen too small, the adaptive scheme keeps subdividing the domain even though the resulting subregions are not significantly lower in rank than their parents. 
This excessive subdivision produces many small patches whose local tensor-train ranks remain essentially unchanged, so that the total number of TT parameters---shown as the solid curve in Figure~\ref{fig:2DGreenOverpatching}(b)---eventually exceeds that of a single full-domain TCI approximation (dotted curve). 
For smaller values of \(\delta\), the function becomes more localized along the Fermi surface, such that overpatching will typically occur only at very small values of \(\chip\).

To mitigate overpatching, several practical safeguards can be adopted. One such strategy is a post-processing merge step: after the recursive phase, identify neighboring patches whose TT cores have identical effective ranks. If, after summing and recompressing the two sibling patches, the resulting bond dimension remains below a prescribed threshold $\chi_{\text{merge}} \ (\gtrsim \chip)$, they can be replaced by the single merged patch.
Such mechanisms would help ensure that domain refinement is guided by the actual approximation error rather than by the arbitrary attainment of the bond-cap, thereby preventing an uncontrolled proliferation of patches and preserving the intended computational savings of patched QTCI.

\subsection{Patch ordering}\label{sec:patch-ordering}

In the adaptive patching scheme, the patch ordering denotes the order in which tensor indices are fixed during the iterative patching of the original tensor. As discussed in Sec.~\ref{sec:adaptivePatchingAlgorithm}, this ordering can be chosen arbitrarily.

\begin{figure}[htbp]
    \centering
    \includegraphics{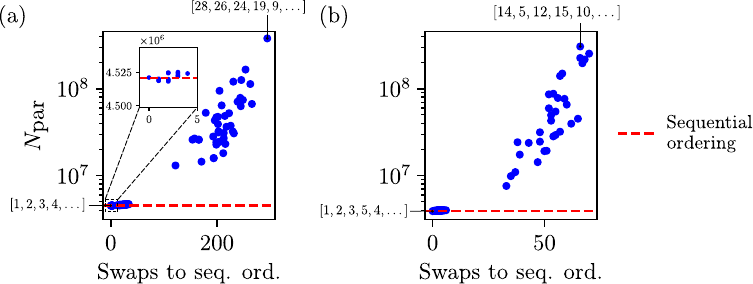}
    \caption{Parameter counts for the patched approximation of \eqref{eq:2DGreen} with $\delta = 10^{-3}$, $\omega = 10^{-1}$, $R = 15$, and $\tau = 10^{-3}$ are shown for (a) interleaved and (b) fused representations. The vertical axis reports the number of parameters, while the horizontal axis indicates the number of adjacent swaps required to recover the sequential ordering $[1,2,3,\dots]$. Each point corresponds to a different patch ordering, with selected orderings explicitly labeled. The red horizontal line denotes the memory usage associated with the sequential ordering. In panel (a), a zoomed-in view of the first few points highlights that the sequential ordering is not always the optimal one for minimizing memory requirements.}
    \label{fig:MemoryvsPordering}
\end{figure}

Figure~\ref{fig:MemoryvsPordering} shows the parameter count of the patched approximation of \eqref{eq:2DGreen} as a function of the number of neighboring swaps required to recover the sequential ordering $[1,2,3,\dots]$ from randomly generated patch orderings used for the approximation, both for interleaved (a) and fused (b) representation (see the caption for the parameter details).

Although the sequential ordering is not always optimal ---especially in the interleaved case (see Fig.~\ref{fig:MemoryvsPordering}(a))---  a clear trend is observed: increasing deviations from the sequential ordering generally lead to a degradation of the patched approximation, reflected in an exponentially larger parameter count. Correspondingly, tests on \eqref{eq:2DGreen} indicate that the optimal patch ordering differs from the sequential one only slightly, in terms of neighboring swaps.

Motivated by these observations, we propose a heuristic algorithm that dynamically identifies a near-optimal patch ordering by exploiting pivots and pivot matrices within the pQTCI framework~\cite{Fernandez2024}. The algorithm is detailed in App.~\ref{app:patchordering}.
In essence, it targets the bond with the largest bond dimension and greedily chooses the next splitting/projection site \(\ell^\star\) that is expected to reduce this bottleneck rank the most.
This is done by using the pivot information to estimate the numerical ranks of the modified pivot matrices obtained by fixing \(\sigma_\ell=v\), and selecting the \(\ell\) that minimizes the resulting rank cost (see Algorithm~\ref{alg:choose-site-with-pivots}).

\section{Patched MPO--MPO contractions}\label{sec:patched-mpo-mpo-contractions}
MPO--MPO contractions are typically the most demanding operations in MPS-based computations: their arithmetic cost scales as \(\mathcal{O}(\chi^{4})\), where \(\chi\) is the bond dimension of the two factors \cite{Stoudenmire2010}.
Such a strong dependence quickly turns the contraction step into a major bottleneck, e.g., in solving the Bethe-Salpeter equation~\cite{Rohshap2025a}. 
Because the bond dimension is the dominant cost driver, a \emph{distributed} strategy that caps the \emph{local} bond dimension---mirroring the philosophy of the patching scheme---promises substantial benefit.
In this section, we consider MPOs of the form in Eq.~\eqref{eq:MPO}.
An approximate contraction of two MPOs is illustrated in Fig.~\ref{fig:MPOMPOcontr}.

In what follows we introduce a \emph{patched MPO--MPO contraction} strategy, and show how it can accelerate several representative tensor contractions encountered in practical applications, already when no adaptivity is introduced and even more so with adaptivity. We start by defining \textit{patched MPOs} and by illustrating how to contract them.

\subsection{Patched Matrix Product Operators (MPO)}

\begin{figure}[htbp]
    \centering \includegraphics{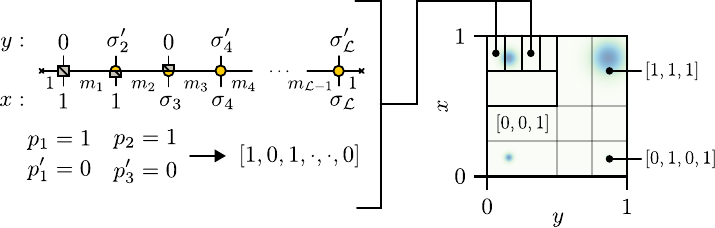}
    \caption{Illustration of a patched MPO. The quantics indices $\bsigma$ and $\bsigma'$ correspond to $x$ and $y$ directions, respectively. The MPO is projected onto a patch by fixing some bits of quantics indices $\boldsymbol{\sigma}$ and $\bsigma'$ to specific values. Additional patches are labeled, assuming an interleaved representation of the $\bsigma$ and $\bsigma'$ indices.
}
    \label{fig:patchMPO}
\end{figure}

Following Sec.~\ref{sec:patching-tensor}, we introduce the notion of a patched MPO, which is an MPO that has been projected on a patch.
We project a site tensor of an MPO on a patch as follows:
\begin{align}
        \raisebox{-0.6cm}{\includegraphics{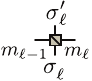}} &= \begin{cases}
		[\mathds{1}]_{m_{\ell-1},m_{\ell}} & \text{if } \sigma_\ell = p_\ell \wedge \sigma_{\ell}^\prime = p_{\ell}^\prime\\
		[0]_{m_{\ell-1},m_{\ell}} & \text{otherwise},
        \end{cases} \label{eq:patchedMPO_twoside}\\[6pt]
        \raisebox{-0.6cm}{\includegraphics{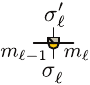}} &= \begin{cases}
	[M_\ell]^{\sigma_{\ell}, p'_\ell}_{m_{\ell-1},m_{\ell}} & \text{if }\sigma_\ell^\prime = p_\ell^\prime\\
		[0]_{m_{\ell-1},m_{\ell}} & \text{otherwise},
        \end{cases} \label{eq:patchedMPO_upside}\\[6pt]
        \raisebox{-0.6cm}{\includegraphics{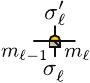}} &= \begin{cases}
	[M_\ell]^{p_\ell,\sigma'_{\ell}}_{m_{\ell-1},m_{\ell}} & \text{if } \sigma_{\ell} = p_{\ell} \\
		[0]_{m_{\ell-1},m_{\ell}} & \text{otherwise}.
        \end{cases} \label{eq:patchedMPO_downside}
\end{align}
A patched MPO can be interpreted through the quantics discretization of a two-dimensional function, where \(f(x(\boldsymbol{\sigma}),y(\boldsymbol{\sigma}')) = F_{\boldsymbol{\sigma}\boldsymbol{\sigma}'} \approx \widetilde{F}_{\boldsymbol{\sigma}\boldsymbol{\sigma}'}\).
By fixing a subset of the binary indices $\bsigma, \bsigma'$, the MPO approximation is restricted to a corresponding subdomain of the function. Figure~\ref{fig:patchMPO} illustrates this idea: some legs of the MPO are fixed, yielding an operator that represents the function only on selected subdomains.

\subsection{Multiplications of patched MPOs}\label{sec:patchedMPOMPOContr}
We now describe how to contract two patched MPOs patch-wise.
Suppose we have two patched MPOs $\widetilde{A}_{\bsigma\bsigma'}$ and $\widetilde{B}_{\bsigma'\bsigma''}$ and contract them as a matrix multiplication:
\begin{equation}
    \widetilde{C}_{\bsigma\bsigma''} = \sum_{\bsigma'} \widetilde{A}_{\bsigma\bsigma'} \widetilde{B}_{\bsigma'\bsigma''},
    \label{eq:patchedMPOMPOContr}
\end{equation}
as illustrated in Fig.~\ref{fig:MPOMPOcontr}.
If the two MPOs are decomposed into $N_A$ and $N_B$ patches, respectively, then the full contraction can be assembled patch-wise by contracting pairs of patches from each factor.
In principle, one would need to consider all $N_A \times N_B$ pairs; however, most of these yield zero because the patches are not \emph{compatible}.

\begin{figure}[htbp]
    \centering \includegraphics[width=0.9\textwidth]{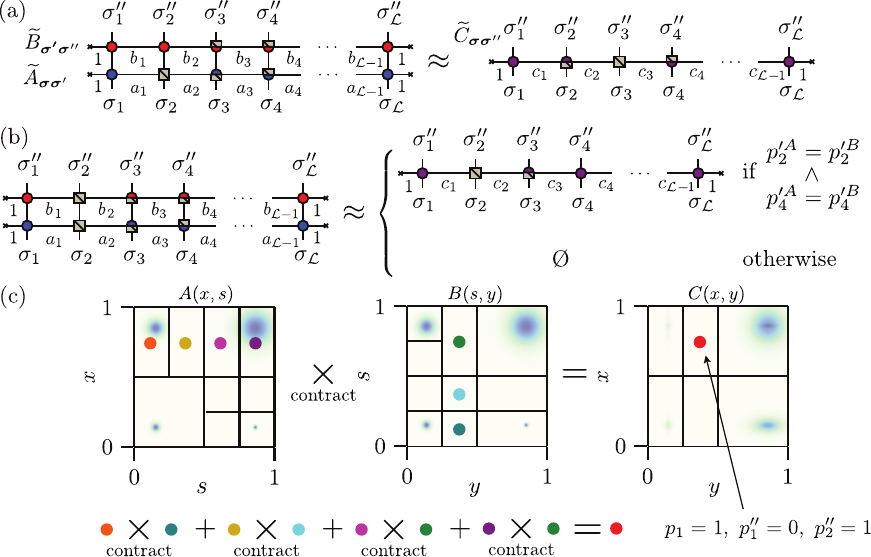}
    \caption{
    Tensor-network diagrams of the patched MPO--MPO contraction. 
    (a) A compatible pair where none of the internal indices (those in $\bsigma'$) are projected simultaneously in both $\widetilde{A}$ and $\widetilde{B}$, always yielding a non-zero contribution.
    (b) A case where both $\widetilde{A}$ and $\widetilde{B}$ project their internal indices at the same sites: the result is non-zero only if the projections match (i.e., $p_2^{\prime A} = p_2^{\prime B}$ and $p_4^{\prime A} = p_4^{\prime B}$), and zero otherwise. 
    (c) Two-dimensional representation of the patched contraction in the QTT case with interleaved quantics indices.
    }
    \label{fig:patchMPOMPOContr}
\end{figure}

Two patches are compatible if and only if their projected internal indices match.
More precisely, for any site $\ell$ where both $\widetilde{A}$ and $\widetilde{B}$ project their internal indices $\sigma'_\ell$, the projections must agree: $p_\ell^{\prime A} = p_\ell^{\prime B}$.
In contrast, the external indices (those in $\bsigma$ and $\bsigma''$) place no such restriction, allowing patches with different external projections to contribute to the result.
Figure~\ref{fig:patchMPOMPOContr}(a) illustrates a compatible pair where none of the internal indices are projected simultaneously in both $\widetilde{A}$ and $\widetilde{B}$, yielding a non-zero contribution.
Figure~\ref{fig:patchMPOMPOContr}(b) shows a case where both $\widetilde{A}$ and $\widetilde{B}$ project their internal indices at the same sites: this pair yields a non-zero contribution only if the projections match (i.e., $p_2^{\prime A} = p_2^{\prime B}$ and $p_4^{\prime A} = p_4^{\prime B}$), and zero otherwise.

The resulting patched MPO $\widetilde{C}_{\bsigma\bsigma''}$ inherits the projected external indices from both $\widetilde{A}$ and $\widetilde{B}$: those projected on $\bsigma$ come from $\widetilde{A}$, while those projected on $\bsigma''$ come from $\widetilde{B}$.
Each output patch in $\widetilde{C}_{\bsigma\bsigma''}$ is obtained by summing contributions from all compatible patch pairs in $\widetilde{A}$ and $\widetilde{B}$.

As a special case, consider the QTT representation where patches are constructed from the coarsest scales, i.e., projections begin at $\ell=1, 2, \ldots$.
In this case, the patched MPO--MPO contraction reduces to a multiplication of two block-structured matrices, as depicted in Fig.~\ref{fig:patchMPOMPOContr}(c). In this particular example, the analogy with standard matrix multiplication becomes explicit: the “row” patches of 
$\widetilde{A}$ are contracted with the “column” patches of $\widetilde{B}$ according to the compatibility rules, exactly as in a block-structured matrix product. The resulting MPO $\widetilde{C}$ inherits the row and column subdivision of $\widetilde{A}$ and $\widetilde{B}$, respectively.
Another important special case is the element-wise (Hadamard) multiplication, which can be implemented by mapping each site tensor of the original QTT onto a diagonal tensor, ensuring $p_\ell^\prime = p_\ell$ in Eq.~\eqref{eq:patchedMPO_twoside}. 
More generally, the patched contraction framework applies to arbitrary patching schemes and is not restricted to these specific cases.

We note that in our numerical examples of MPO--MPO contraction, we truncate the resulting TT in each patch using the Frobenius norm with a local error measure, i.e., the error is measured relative to the norm of the TT within each patch.
As discussed in Sec.~\ref{sec:adaptivePatchingAlgorithm}, using a local error measure could lead to a large increase of the bond dimension for patches where the function takes small values.
Although we did not observe this behavior in our numerical examples, this remains a potential issue that should be addressed in future work.

\subsection{Adaptive patched MPO--MPO contraction}
\label{sec:AdaptivePatchMPOMPOContr}
In the previous section, the patch layout is fixed by the two input MPOs, namely by how their external indices are projected.
As a consequence, the resulting patch structure is not necessarily well matched to the product of the two MPOs.
We now introduce a simple adaptive scheme that refines the patch structure of the result.

For simplicity, we describe the version that refines indices from the first to the last.
The procedure is as follows:
\begingroup
\renewcommand{\labelenumi}{(\arabic{enumi})}
\begin{enumerate}
    \item Fix a cap on the local bond dimension \(\chip\) and a target accuracy \(\tau\).
    \item Perform a patched MPO--MPO contraction as in the previous section, capping the bond dimension of every intermediate and output patch at \(\chip\) while accumulating contributions.
    \item Mark as unconverged any output patch whose rank reaches \(\chip\), and record all patches in \(\widetilde{A}\) and \(\widetilde{B}\) that contributed to these unconverged outputs.
    \item Refine those contributing patches of \(\widetilde{A}\) and \(\widetilde{B}\) by splitting them according to the first external index that has not yet been patched.
    \item Recompute only the unconverged output patches of \(\widetilde{C}\) with the refined inputs, and repeat until all output patches are converged, i.e., their ranks remain strictly below \(\chip\).
\end{enumerate}
\endgroup

\begin{figure}[tp]
    \centering
    \includegraphics[width=\textwidth]{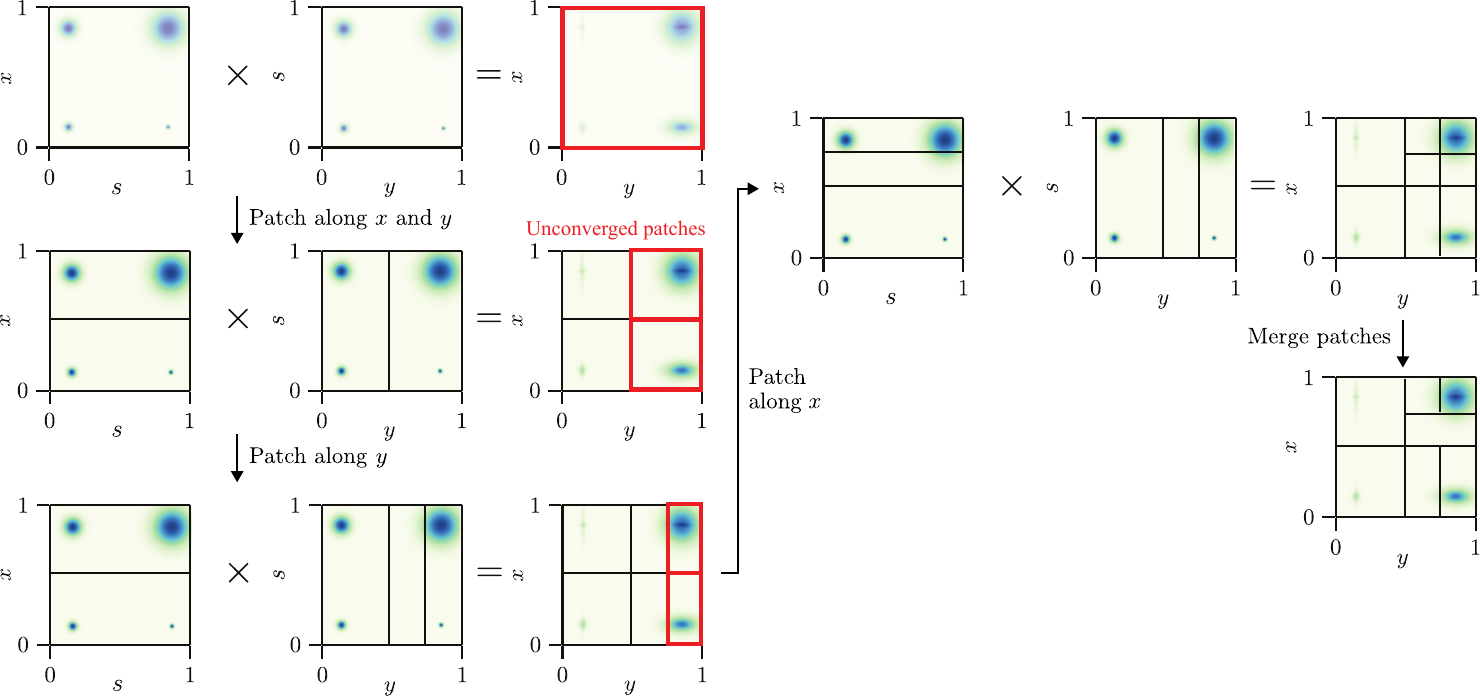}
    \caption{
        Illustration of the adaptive patched MPO--MPO contraction. We start with contracting two MPOs (matrices) and patch the external indices adaptively.
        The output patches highlighted in red are the ones that are not converged.
        We can patch both of the external indices for $x$ and $y$ directions simultaneously, or patch one of the external indices for $x$ direction and the other for $y$ direction.
    }
    \label{fig:adaptivePatchContr}
\end{figure}

We illustrate the algorithm in Figure~\ref{fig:adaptivePatchContr} using the sample example in Fig.~\ref{fig:patchMPOMPOContr}(c).
This refinement produces smaller output patches in \(\widetilde{C}\) exactly where needed, which makes convergence more likely.
Once all patches have converged, some may have bond dimensions well below \(\chip\).
In that case we attempt to merge \textit{neighboring} patches by summing their TTs; a merge is accepted if the resulting bond dimension stays below \(\chip\).
Here, \textit{neighboring} patches are those that have the same projected external indices except for the last projected index.

The same strategy also applies when the inputs are unpatched (i.e., \(\widetilde{A}\) and \(\widetilde{B}\) are single patches with bond dimensions much larger than \(\chip\)).
Although the algorithm might appear more costly than a single contraction, in practice the overhead is negligible: as soon as an output patch reaches the bond-cap \(\chip\) we can stop the corresponding local contraction early, for example by monitoring the SVD rank in the zip-up sweep.
Consequently, each attempt is inexpensive thanks to the tight rank bound \(\chip\).

Since we refine patches adaptively, it is crucial to use a global error measure for truncation (see Secs.~\ref{sec:adaptivePatchingAlgorithm} and \ref{sec:patchedMPOMPOContr}) to avoid blowing up the bond dimension for patches where the function takes small values.
In practice, when using SVD, we can truncate the TT on each patch $p$ according to $\bigl\|\tilde{C}^p_\bsigma - C^p_\bsigma\bigr\|_\mathrm{F} < \tau' \bigl\| C_\bsigma \bigr\|_\mathrm{F}$, where $\tau'~(\le \tau)$ is a temporary tolerance and $C_\bsigma$ is the original tensor.
The tolerance $\tau'$ is adjusted so that the total error $\sum_{p=1}^{\Np} \bigl\|\tilde{C}^p_\bsigma - C^p_\bsigma\bigr\|_\mathrm{F}$ remains below $\tau \bigl\| C_\bsigma \bigr\|_\mathrm{F}$, which may typically require $\tau' \sim \tau / \Np$.

\subsection{Estimates on computational complexity}\label{sec:complexity-elementwise}
A precise complexity estimate of patched MPO--MPO contractions is difficult, as it depends on the structure of the input MPOs and on how patching reduces their bond dimensions.
Nevertheless, we illustrate how patching can reduce the contraction cost.
Consider contracting two ``matrix-like'' MPOs, $\widetilde{A}(x,s)$ and $\widetilde{B}(s,y)$, patched only along one internal direction (columns for \(\widetilde{A}\), rows for \(\widetilde{B}\)), as in Fig.~\ref{fig:adaptivePatchContr}.

We assume that the bond dimension of the unpatched input MPOs is $\chi$ and that the bond dimension of each input patch scales as $\chi/\sqrt{\Np}$, so that the total number of parameters is preserved.
This scaling is consistent with our numerical observations, as shown in Fig.~\ref{fig:patchEvolution+AlgTree}.
We also assume that the bond dimension of each output patch is $\chi/\sqrt{\Np}$.
Under these assumptions, the total contraction cost scales as \(\mathcal{O}\bigl(\mL\, N_\mathrm{pair}\, (\chi/\sqrt{\Np})^4\bigr)\), where
\(N_\mathrm{pair}\) is the number of compatible pairs of patches.
Because we patch only along the column ($\widetilde{A}$) and row ($\widetilde{B}$) indices, only one compatible pair contributes to each output patch, hence \(N_\mathrm{pair} = \Np\).
Therefore the overall cost scales as \(\mathcal{O}(\chi^4 \mL/\Np)\).

The same reasoning applies to element-wise multiplication, where both inputs are diagonal (e.g., $\widetilde{A}(x,x') = \widetilde{A}(x)\delta_{x, x'}$).
Consider two \emph{unpatched} diagonal MPOs of bond dimension \(\chi\).
We first split each MPO into \(\Np\) patches of maximum bond dimension \(\chip\), chosen such that the total number of parameters is preserved,
\(\Np\,\chip^2 \sim \chi^2\), i.e. \(\chip \sim \chi/\sqrt{\Np}\).
The local element-wise multiplication on a single patch scales as \(\mathcal{O}(\mL\,\chip^4)\), hence the total local cost scales as
\(\mathcal{O}(\mL\,\Np\,\chip^4)=\mathcal{O}(\chi^4\mL/\Np)\).
The overhead of splitting and recombining patches back into a single MPO is dominated by TT recompression during successive merges.
Assuming that disjoint neighboring patches combine with bond dimensions scaling as \(\chi_{(1,2)}^2\simeq \chi^2_{(1)} + \chi_{(2)}^2\) (so that \(\chi_{(1,2)}\simeq \sqrt{2}\chi_{(1)}\) for $\chi_{(1)}\simeq\chi_{(2)}$),
the total cost of merging two patches scales as \(\mathcal{O}(\mL\,\chip^3)\).
Merging iteratively in a binary merge tree gives
\(\mathcal{O}\bigl(\mL\,\Np\,\chip^3\bigr)=\mathcal{O}\bigl(\mL\,\chi^3/\sqrt{\Np}\bigr)\), which is bounded by \(\mathcal{O}(\mL\,\chi^3)\).
Altogether, the cost of element-wise multiplication scales as \(\mathcal{O}(\mL\,\chi^3 + \chi^4\mL/\Np)\).
In particular, for sufficiently fine patching the speedup saturates and the cost can approach the \(\mathcal{O}(\mL\,\chi^3)\) regime.
This should be contrasted with recent algorithms that target element-wise multiplication directly on (unpatched) tensor trains and achieve \(\mathcal{O}(\chi^3)\) computational complexity with \(\mathcal{O}(\chi^2)\) memory footprint~\cite{Michailidis2025TTM}.
Our patching strategy is complementary: it can exploit locality to reduce work, but may incur additional memory overhead from storing multiple patches and intermediate recompressions.

Why does patching reduce the total cost?
It exploits the locality of the contraction.
This is most transparent for the element-wise multiplication of two functions, $f(x) = f_A(x) f_B(x)$, which corresponds to multiplying two diagonal MPOs.
To evaluate $f(x)$ at a given $x$, one needs only $f_A(x)$ and $f_B(x)$, and does not need $f_A(x')$ or $f_B(x')$ for $x' \neq x$.
Patching makes this locality explicit and avoids unnecessary interactions between unrelated regions.

The computational complexity depends on how the input MPOs are patched.
If the input MPOs are patched along both internal and external indices, yielding square-like patches in the matrix representation, more patch pairs must be contracted, as illustrated in Fig.~\ref{fig:patchMPOMPOContr}, which increases the overall cost.
In this case, $N_\mathrm{pair} = \mathcal{O}(\Np^{3/2})$, leading to an overall cost scaling of \(\mathcal{O}(\chi^4 \mL/\sqrt{\Np})\), which is less favorable than the \(\mathcal{O}(\chi^4 \mL/\Np)\) scaling achieved when patching only along one direction. A detailed analysis illustrating how different patching patterns affect the contraction cost is provided in App.~\ref{app:patchPatternMPOMPOContr}.
 
A promising direction for future work would be to develop methods for repatching MPOs in different orders to optimize the contraction cost.

\section{Example: bare susceptibility}\label{sec:bubble}

Having tested the usefulness of patching in the compression of one-particle Green's functions, let us move to an example where patching helps with the evaluation of Feynman diagrams, i.e. convolutions of Green's functions. The simplest Feynman diagram involving a convolution is the so-called ``bubble'' diagram, consisting of two Green's function lines connected at both ends. This bubble diagram represents the bare susceptibility (``bare'' because it does not include the vertex correction) and is needed in numerous many-body methods~\cite{Aryasetiawan1998,  BickersFLEX,Vilk1997,RevModPhys.90.025003, Rohshap2025a}. A similar diagram is also arises in the computation of optical conductivity.

\subsection{Matsubara axis}
We begin by considering convolutions of imaginary-frequency (Matsubara) propagators, dealing with real-frequency ones in the next subsection. The imaginary-frequency case is simpler, since the discreteness of the Matsubara frequencies, which are separated by $2\pi T$, brings a natural temperature broadening and in consequence lower bond dimensions. We are interested in the bare susceptibility, which is given by the convolution
\begin{equation}
\chi_0 (\bq, i\omega_n) = \frac{1}{(2\pi)^D\beta} \int_\mathrm{BZ} d^D k\, \sum_{i\nu_n} G(\bk, i\nu_n)\, G(\bk + \bq, i\nu_n + i\omega_n),
 \end{equation}
where \(D\) is the dimensionality of the system, $\nu_n$ and $\omega_n$ are fermionic and bosonic Matsubara frequencies, and $\beta=1/T$ is the inverse temperature.
The integral runs over the Brillouin zone, denoted \(\mathrm{BZ}\).

The convolution can be evaluated efficiently using Fourier transforms and the convolution theorem as
\begin{equation}\label{eq:convtheorem_matsubara}
\chi_0 (\bq, i\omega_n) = FT \left[\hat \chi_0(\br, \bar\tau) \right] = FT \left[\hat G(\br,\bar\tau)\, \hat G(-\br,-\bar\tau)\right].
 \end{equation}
We use \(\bar \tau\) for imaginary time to avoid confusion with the truncation tolerance \(\tau\).
We define the forward and inverse Fourier transforms of the Green's function as
\begin{subequations}\label{eq:FTGreensFunc}
\begin{align}
     G(\bk, i\nu_n) &=  \sum_{\br} \int_{0}^{\beta} d\bar\tau\;  e^{i\bk\cdot\br+i\nu_n\bar\tau}  \hat G(\br, \bar\tau) 
     \label{eq:FTGreensFunc_a} 
     \\
    \hat G(\br, \bar\tau) &= \frac{1}{(2\pi)^D\beta} \int_{\mathrm{BZ}} d^Dk\, \sum_{i\nu_n} e^{-i\bk\cdot\br-i\nu_n\bar\tau} G(\bk, i\nu_n).
    \label{eq:invFTGreensFunc_b} 
\end{align}
\end{subequations}
Here and above the sum $\sum_{i\nu_n}$ runs over the Matsubara frequency grid \(\nu_n=(2n+1)\pi/\beta\) with $n=-N_{\nu}/2, -N_{\nu}/2+1, \dots, N_{\nu}/2-1$. 

\begin{figure}[htbp]
    \centering
    \includegraphics{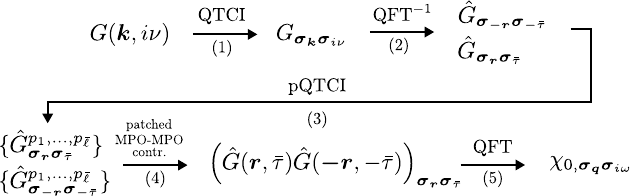}
    \caption{Pipeline for computing the bare susceptibility \(\chi_{0}(\bq,i\omega)\).}
    \label{fig:bubbleFlowchart}
\end{figure}

\begin{figure}[htbp]
    \centering 
    \includegraphics{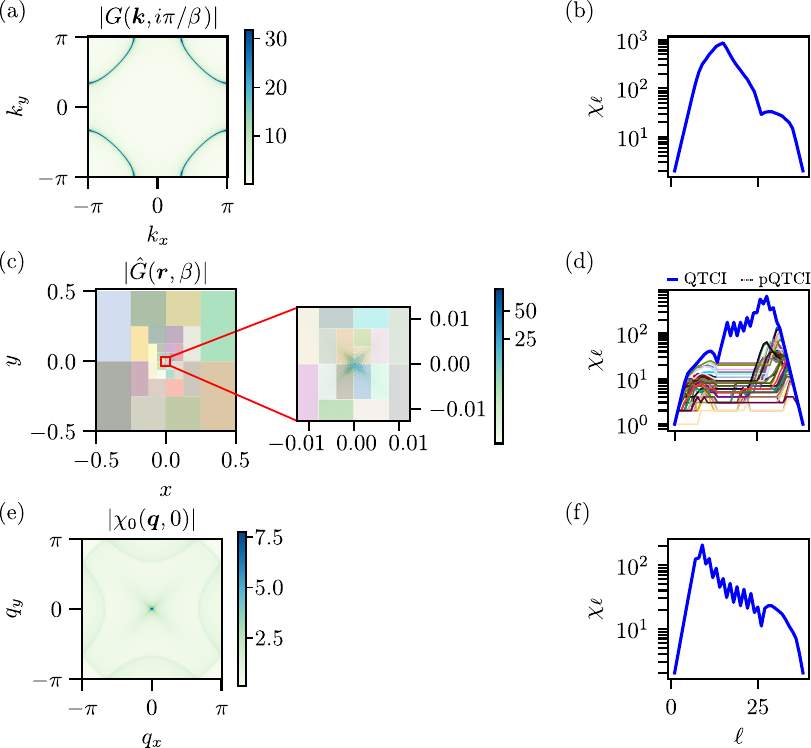}
    \caption{
    Heatmap and bond dimensions overview of the bare-susceptibility workflow with $\mR=13$, $\beta=100$, $\mu=1$ and tolerance $\tau=10^{-5}$. Left column: representative slices of each object. Right column: bond dimension profiles of the corresponding full objects.
    (a) Absolute value of $G(\bk,i\pi/\beta)$ at the first positive fermionic Matsubara frequency.
    (b) Bond dimension profile of the TCI approximation of $G(\bk,i\nu_n)$.
    (c) $\hat{G}(\br,\beta)$ in real space at imaginary time $\bar{\tau}=\beta$.
    (d) Bond dimension profiles of $\hat{G}(\br,\bar{\tau})$ for the unpatched and patched (pQTCI) approximations; the bond-dimension cap is set to $\chi_p=181$.
    (e) Bare susceptibility $\chi_{0}(\bq,0)$ at the zeroth bosonic Matsubara frequency.
    (f) Bond dimension profile of the QTT approximation of $\chi_{0}(\bq,i\omega_n)$. }
    \label{fig:heatmapBubble}
\end{figure}

\begin{figure}[tp]
    \centering
    \includegraphics{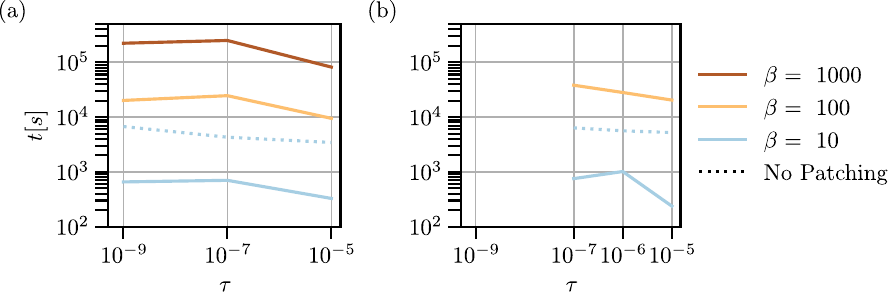}
    \caption{CPU time for the patched element-wise product compared with the monolithic contraction as a function of tolerance \(\tau\) and at different inverse temperatures \(\beta\) for $\mu=0$ (a) and $\mu=1$ (b). For each set of parameters the optimal bond-dimension caps $\chip$ are chosen (cf. Figs.~\ref{fig:memoryTime2DGreen}, \ref{fig:deltavsMemoryTime}). The benchmark was performed on a single Intel\textsuperscript{\textregistered} Xeon\textsuperscript{\textregistered} E5-2680 v4 @ 2.40 Ghz.}
    \label{fig:bubbleResults}
\end{figure}

We consider a 2D square lattice with the bare Green's function presented in~\eqref{eq:2DGreen}, \(G(\bk, i\nu_n) = (i\nu_n + \mu - \varepsilon_\bk)^{-1}\), which we QTT-compress jointly over momentum and Matsubara frequency, as the starting point for evaluating \(\chi_{0}(\bq,i\omega_n)\). The complete workflow is depicted in the flowchart of Figure~\ref{fig:bubbleFlowchart}. The index ordering for the QTTs is chosen as follows: in momentum space, the $k_x$ and $k_y$ quantics legs are interleaved and the Matsubara frequency legs are appended at the end; in real space, the imaginary time quantics legs are placed at the beginning of the TT and the $x$ and $y$ legs are interleaved thereafter. Representative frequency and imaginary-time slices of $G(\bk,i\nu_n)$, $\hat{G}(\br,\bar{\tau})$ and $\chi_0(\bq,i\omega_n)$ are shown as heatmaps in Fig.~\ref{fig:heatmapBubble}(a),(c),(e), with the bond dimension profiles of the corresponding full objects shown in  Fig.~\ref{fig:heatmapBubble}(b),(d),(f). The sharp arc-like structures present in $G(\bk,i\pi/\beta)$ at $\beta =100$ (cf. Fig.~\ref{fig:heatmapBubble} (a)) reflect the shape of the Fermi surface in the doped 2D Hubbard model, which is almost incompressible (cf. large bond dimensions in Fig.~\ref{fig:heatmapBubble} (b)). The sharpness of the features, and hence the bond dimension, increases with inverse temperature $\beta$; see also Fig.~\ref{fig:2DGreenErrorHeatmap} where the broadening $\delta$ plays the same role as temperature here. After patching — which is applied exclusively to the real space quantics indices of \(\hat G(\br,\bar{\tau})\) (i.e.\ starting from the $\mR$th leg), as evidenced by the flat intermediate regions in the bond dimension profile in Fig.~\ref{fig:heatmapBubble}(d) — the bond dimension per patch is significantly lower. Remarkably, this reduces the runtime of the bubble calculation by an order of magnitude relative to the unpatched computation, as shown in Fig.~\ref{fig:bubbleResults}(b). For $\mu=0$, the Fermi surface is a square (not shown) leading to slightly lower bond dimensions and shorter runtimes both for the patched and unpatched implementations (cf. Fig.~\ref{fig:bubbleResults}(a)), compared to the $\mu=1$ case. 

\subsection{Real frequency axis}

On the real-frequency axis, the momentum convolution remains unchanged, but the discrete Matsubara sum becomes a continuous frequency integral.
We focus on the retarded bubble, which in spectral form can be written as
\begin{equation}\label{eq:bubble_real}
\chi^\mathrm{R}_0 (\bq, \omega) = \frac{1}{(2\pi)^D} \int_\mathrm{BZ} d^D k\, \int_{-\infty}^\infty d\nu\; n_\mathrm{F}(\nu)\, A(\bk, \nu)
 \Big[G^{\mathrm{R}}(\bk + \bq, \nu + \omega) + G^{\mathrm{A}}(\bk - \bq, \nu - \omega) \Big].
\end{equation}
Here $G^{\mathrm{R}}$ is the retarded Green's function (we use the model~\eqref{eq:2DGreen} with $\delta>0$), $G^{\mathrm{A}}(\bk, \nu) = [G^{\mathrm{R}}(\bk, \nu)]^*$ is the advanced Green's function, $A(\bk, \nu) = -\frac{1}{\pi} \operatorname{Im} G^{\mathrm{R}}(\bk, \nu)$ is the spectral function, and $n_\mathrm{F}(\nu)=\bigl(1+e^{\beta\nu}\bigr)^{-1}$ is the Fermi--Dirac distribution.
The integrand features Lorentzian peaks of width $\sim\delta$, which can be difficult to resolve for small $\delta$.

In practice, we evaluate~\eqref{eq:bubble_real} using the convolution theorem by Fourier transforming to real space and real time.
Analogous to~\eqref{eq:convtheorem_matsubara}, the retarded bubble in $(\br,t)$ space can be expressed as
\begin{equation}
\hat \chi^\mathrm{R}_0(\br, t) = 2\,\operatorname{Im} \bigl[\hat G^<(-\br, -t)\, \hat G^\mathrm{R}(\br, t)\bigr],
\end{equation}
where $G^<(\bk, \nu) = 2\pi i\, n_\mathrm{F}(\nu) A(\bk, \nu)$ and $\hat G^<$ denotes its Fourier transform.
This reduces the computationally intensive part of the workflow to an element-wise multiplication in $(\br,t)$ space, followed by an inverse Fourier transform back to $(\bq,\omega)$.

This is exactly the setting where patching is most useful: we patch both QTT factors representing $\hat G^<(\br,t)$ and $\hat G^{\mathrm{R}}(\br,t)$, multiply patch-wise to cap intermediate bond dimensions, and finally unpatch (sum and recompress) to obtain a single QTT representation of the product.

\begin{figure}[htbp]
    \centering
    \includegraphics{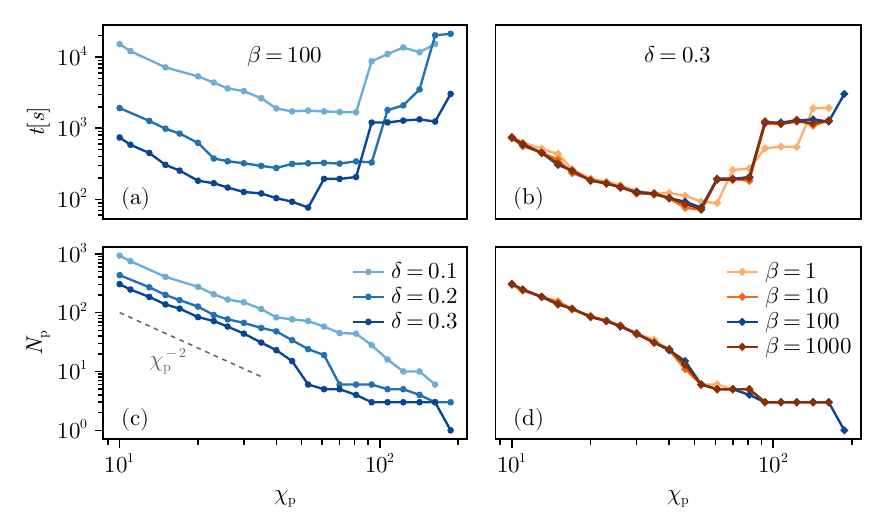}
    \caption{Scaling behavior for the 2D real-frequency bubble computation.
    We use interleaved index ordering.
    Patching proceeds from coarse to fine quantics digits in both time and space (i.e., from most significant bits to least significant bits).
    We use $\mR_\nu = 12$ frequency bits, $\mR_k = 8$ momentum bits per spatial dimension (28 bits total), tolerance $\tau = 10^{-3}$, and chemical potential $\mu = 1$.
    The left column shows results for varying broadening $\delta$ at fixed inverse temperature $\beta = 100$; the right column shows results for varying $\beta$ at fixed $\delta = 0.3$. Note that the $\delta = 0.3$, $\beta = 100$ curve (dark blue) appears in both columns.
    (a,\,b)~Total wall time for the patch--multiply--unpatch procedure. An AMD Ryzen 9 7900X CPU was used to perform the benchmark.
    (c,\,d)~Number of patches $N_{\mathrm{p}}$ in the product $\hat G^<(-\br, -t)\,\allowbreak \hat G^{\mathrm{R}}(\br, t)$ that forms the bubble integrand, i.e., after multiplication and before unpatching.
    The dashed line in (c) shows $\chi_{\mathrm{p}}^{-2}$ scaling for reference.}
    \label{fig:realfreqbubbleresults}
\end{figure}

Figure~\ref{fig:realfreqbubbleresults} benchmarks the numerical costs of this \emph{patch--multiply--unpatch} kernel in isolation (excluding the TCI and QFT steps), since it dominates runtime and memory usage once the two input objects have been constructed.
Panels (a) and (b) show the total wall time as a function of the bond-cap $\chi_{\mathrm{p}}$ for varying broadening $\delta$ at fixed $\beta$ and for varying $\beta$ at fixed $\delta$, respectively.
In all cases we observe a pronounced optimum: for large $\chi_{\mathrm{p}}$ the patch-wise products develop large intermediate bond dimensions (up to $\sim\chi_{\mathrm{p}}^{2}$ before truncation), which increases both the memory footprint and the cost of the subsequent TT rounding step, whereas for small $\chi_{\mathrm{p}}$ the increased number of patches (and corresponding bookkeeping and merging overhead) dominates the runtime.
On the 128~GB machine used for these benchmarks, attempts to increase $\chi_{\mathrm{p}}$ beyond the values shown for $\delta=0.2$ and $0.3$ exhausted available memory.

Panels (c) and (d) report the corresponding number of patches $N_{\mathrm{p}}$ in the product $\hat G^<(-\br, -t)\,\allowbreak \hat G^{\mathrm{R}}(\br, t)$ prior to unpatching.
The near-independence of $N_{\mathrm{p}}$ on $\beta$ reflects that temperature enters only through the smooth cutoff in $n_\mathrm{F}(\nu)$; in the present parameter regime, the dominant source of localization (and thus patch refinement) is controlled by the broadening $\delta$.
Consistent with the discussion in Sec.~\ref{sec:adaptivePatchingAlgorithm}, we find approximately $N_{\mathrm{p}}\propto \chi_{\mathrm{p}}^{-2}$.

\section{\label{sec:BSE}Bethe-Salpeter equation}

The Bethe-Salpeter equation (BSE) is an analogue of the Dyson equation and represents a resummation of a particular class of two-particle Feynman diagrams. It is the central equation and the numerical bottleneck in numerous many-body Green's function approaches~\cite{RevModPhys.90.025003}. In an earlier work~\cite{Rohshap2025a} some of us applied the QTT representation to solve a self-consistent set of quantum field theory equations (parquet equations) connecting different two-particle scattering amplitudes (vertices). This approach requires evaluating the BSE at each iteration  and it dominates the computational effort of the entire self-consistent scheme. 

To show the application of our patching scheme to the BSE, we restrict attention to a relatively simple model, in which the vertices have only energy (frequency) dependence of the scattering particles (in more complicated models, e.g. the Hubbard model, they also depend on momenta). Concretely,  we will focus on the reduction of the Hubbard model on a lattice to a single isolated site whose Hamiltonian is
\begin{equation}
    \mathcal{H} = U\hat{n}_{\uparrow}\hat{n}_{\downarrow} - \mu(\hat{n}_{\uparrow} - \hat{n}_{\downarrow}),
\end{equation}
where the hopping between sites is set to zero and $U$ is the onsite repulsion and $\mu$ the chemical potential. We set here $\mu= U/2$, which means that our site is singly occupied  (half-filled). The operators  $\hat{n}_{\uparrow(\downarrow)}$ are number operators counting the number of electrons with spin ${\uparrow(\downarrow)}$. Despite its simplicity, this \emph{atomic limit} reproduces many key features of the Hubbard model in the strong-coupling regime \cite{Thunstrom2018} and has the advantage that the analytical expressions for the two-particle vertices are known \cite{Thunstrom2018}.


The BSE connects the irreducible vertex, which we here denote as $\Gamma$ in analogy with Ref.~\cite{Rohshap2025a}, and the full vertex $F$. For the density $(d)$  component (one of the spin combinations of incoming and outgoing particles in the two-particle scattering) of the vertex it reads
\begin{align}
    F_{d}^{\nu\nu'\omega} &= \Gamma_d^{\nu\nu'\omega} - \frac{1}{\beta^2} \sum_{\nu_1\nu_2} \Gamma_d^{\nu\nu_1\omega}\chi_{0,ph}^{\nu_1\nu_2\omega}F_d^{\nu_2\nu'\omega},
    \label{eq:bothBSEs}
\end{align} 
where $\nu,\nu',\nu_1,\nu_2$ (and $\omega$) are fermionic (and bosonic) discrete Matsubara frequencies and  $\chi_{0,ph}^{\nu_1\nu_2\omega}$ is a product of Green's functions (see Ref.~\cite{Rohshap2025a} for details), which is diagonal in the $\nu_1, \nu_2$ indices, and $\beta$ is the inverse temperature. 

The right-hand side of Eq.~\eqref{eq:bothBSEs} is a three-indexed three-tensor contraction that must be evaluated many times during the iterative parquet loop, and thus dominates the overall cost.
In Ref.~\cite{Rohshap2025a}, equation~\eqref{eq:bothBSEs} was evaluated using QTCI-based tensor trains and the standard MPO--MPO contraction toolkit.
Here, we revisit the same contraction but replace the monolithic MPO--MPO multiplication by the \emph{patched contraction} strategy, expecting improved efficiency from the control of local bond dimensions of the factor MPOs.
In practice, we compress each vertex as a \emph{single} three-frequency quantics tensor train (a three-dimensional MPS over $(\nu,\nu',\omega)$); fixed-$\omega$ slices are used only for visualization.

\begin{figure}[htpb]
    \centering
    \includegraphics{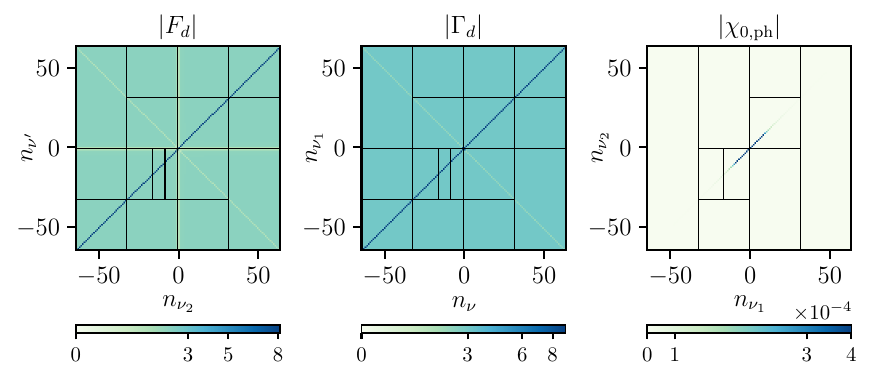}
    \caption{pQTCI compression of the vertices in the Bethe-Salpeter equation at $\omega=0$, $U=3.0$ and $\beta=1.0$. Panels show $|F_d|$ , $|\Gamma_d|$  and $|\chi_{0,ph}|$ (from left to right) on the respective $(\nu_2,\nu')$, $(\nu_1,\nu)$ and $(\nu_2,\nu_1)$ grids for $\mR=7$ and tolerance $\tau=10^{-7}$. Patch boundaries reveal how the adaptive slicing refines only those regions where the vertex is more interesting, also for three-dimensional objects.
    }
    \label{fig:patchedVertices}
\end{figure}

\begin{figure}[ht!]
    \centering
\includegraphics{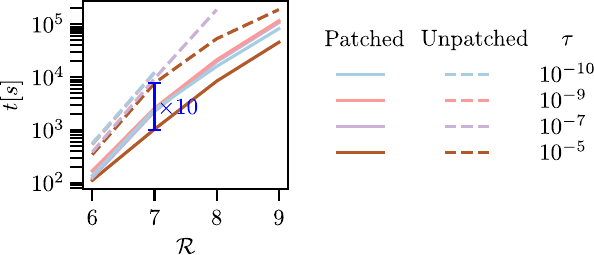}
    \caption{CPU time for the BSE vertex contraction versus the frequency resolution
    \(\mR\) (bits per Matsubara axis).
    Solid curves: patched MPO--MPO contraction;
    dashed curves: conventional (non-patched) contraction. Colors denote the compression tolerance \(\tau\).
    }
    \label{fig:BSEresult}
\end{figure}

The initial stage of the patched BSE scheme is visualized in Figure~\ref{fig:patchedVertices}.
Using the interleaved slicing order, each two-particle vertex is compressed with pQTCI so that the resulting patch grid adapts to the structure of the data.
For clarity we display two-dimensional cuts at the bosonic frequency $\omega=0$; the axes correspond to the fermionic frequencies on a $2^{\mR}\times2^{\mR}$ mesh with $\mR = 7$.
The color scale shows $|F_d|$, $|\Gamma_d|$ and $|\chi_{0,ph}|$, respectively, each reconstructed from their patched approximation up to a tolerance $\tau=10^{-7}$. One sees that smaller tiles concentrate in the regions where the vertices exhibit pronounced structure, whereas featureless areas are covered by larger patches. Before performing the contraction, each TT (patch) obtained from the vertex approximation is converted into a patched MPO, following the procedure described in Ref.~\cite{Rohshap2025a} (cf. Fig.~8). 
Figure~\ref{fig:BSEresult} compares the wall-clock time required to evaluate the contraction on the right-hand side of the BSEs in Eq.~\eqref{eq:bothBSEs} with and without patching.  
Calculations were performed on an
Intel Xeon E5-2680 v4 @ 2.40 GHz; the horizontal axis shows the number of bits \(\mR\) used per fermionic frequency (i.e.\ \(\mR=\log_{2}N_{\nu}\) with $N_{\nu}$ total frequencies). Four target tolerances \(\tau\) are reported.  
Over the entire range the patched algorithm outperforms the conventional single-MPO contraction by up to an order of magnitude.  
For the stringent tolerances \(\tau=10^{-9}, 10^{-10}\) the monolithic approach was no longer feasible beyond \(\mR=7\) owing to excessive memory requirements, whereas the patched routine remained tractable.

\section{Conclusion}
\label{sec:conclusion}

We have introduced an adaptive patching strategy for Quantics Tensor Trains that exploits block-sparse tensor-core structures to reduce bond dimensions and computational costs in high-dimensional problems.
By combining this idea with quantics Tensor Cross Interpolation (pQTCI) and patched MPO--MPO contractions, we demonstrated substantial savings in memory and runtime for sharply localized functions, as exemplified by two-dimensional Green's functions, bubble diagrams, and Bethe-Salpeter equations.
These results show that adaptive partitioning can focus numerical effort on regions where the underlying functions exhibit separable structures at some length scales.

The proposed method can be regarded as a combination of conventional divide-and-conquer strategies, such as adaptive mesh refinement (AMR), with state-of-the-art tensor network methods.
Conventional AMR is a powerful technique but refines the grid only along a single direction in scale space, typically from coarse to fine resolution (or vice versa)~\cite{Zhang2021-qp}.
In contrast, the present approach is more general, since partitioning can be performed at intermediate scales in arbitrary orders.
It would be desirable to develop heuristic algorithms that automatically determine near-optimal partitioning orders. We present a first step in that direction in App.~\ref{app:patchordering}.

We leave the application of our patching scheme to a wider range of problems as an interesting prospect for future work.
In particular, fast element-wise function operations are  widely recognized as a key challenge for tensor-train based simulations, since they can dominate both runtime and memory usage; see Sec.~\ref{sec:complexity-elementwise} and Refs.~\cite{Michailidis2025TTM,meng_recursive_2026,ritter_elementwise_2026}.
For instance, the MPO--MPO contraction is also a critical bottleneck in computing the thermal density matrix at finite temperatures in XTRG~\cite{XTRGthermal}.
It is an intriguing question whether the thermal density matrix exhibits a similar quasi-block structure, such that its contractions could be further accelerated by the adaptive patching techniques proposed here.
Moreover, AMR-based methods are central to relativistic field-theory simulations in numerical relativity and cosmology~\cite{Andrade2021-ws,Drew2022-qg,Drew2023-oh}, suggesting that related adaptive patching ideas may eventually help reduce the cost of evolving high-dimensional field configurations in those settings.
Finally, the patching scheme could be further enhanced by exploiting symmetries of the underlying problem: symmetric functions produce patches related by symmetry transformations, so that only an irreducible subset of patches would need to be computed explicitly, with the remainder reconstructed by symmetry.

\section*{Acknowledgements}
We thank Markus Frankenbach, Simone Foder\`a, Yuehaw Khoo, Ken'ichi Saikawa, and Miles Stoudenmire for valuable discussions. 
We acknowledge the use of large language models (LLMs) for assistance in the preparation of this manuscript, particularly for text editing and language refinement. The core scientific content, methodology, and results remain entirely our own work.
The Flatiron Institute is a division of Simons Foundation.


\paragraph{Software} The code used for this work is publicly available through the tensor4all collaboration\cite{tensor4all.org}. All results presented in this manuscript were produced using the \texttt{TCIAlgorithms.jl} and \texttt{PartitionedMPSs.jl} libraries, where the former depends on \texttt{TensorCrossInterpolation.jl}.

\paragraph{Funding information}

This work was funded in part by the Deutsche Forschungsgemeinschaft under Germany’s Excellence Strategy EXC-2111 (Project No. 390814868). It is part of the Munich Quantum Valley, supported by the Bavarian state government with funds from the Hightech Agenda Bayern Plus. We further acknowledge support from DFG Grant No. LE 3883/2-2. We gratefully acknowledge computational resources from Grant No. INST 86/1885-1 FUGG of the German Research Foundation (DFG) and from the GCS Supercomputer SuperMUC-NG at the Leibniz Supercomputing Centre in Munich provided
by the Gauss Centre for Supercomputing e.V.
H.S. was supported by JSPS KAKENHI Grant Nos. 22KK0226 and 23H03817, JSPS Bilateral Program No. JPJSBP120252002, and JST FOREST Grant No. JPMJFR2232, Japan. This research was funded in part by the Austrian Science Fund (FWF) Grant DOI 10.55776/P36332, 10.55776/V1018, and 10.55776/PIN4372024.

\begin{appendix}

\section{Patched Quantics Tensor Cross Interpolation (pQTCI)}\label{app:pqtci}

    The adaptive patching algorithm generates a set of patches according to the structure of the tensor to be approximated.
    It can be combined with any scheme that generates TT approximations of tensors; here, we describe \emph{patched Quantics Tensor Cross Interpolation} (pQTCI), its combination with Tensor Cross Interpolation.
    
\begin{figure}[htbp]
        \centering
        \includegraphics{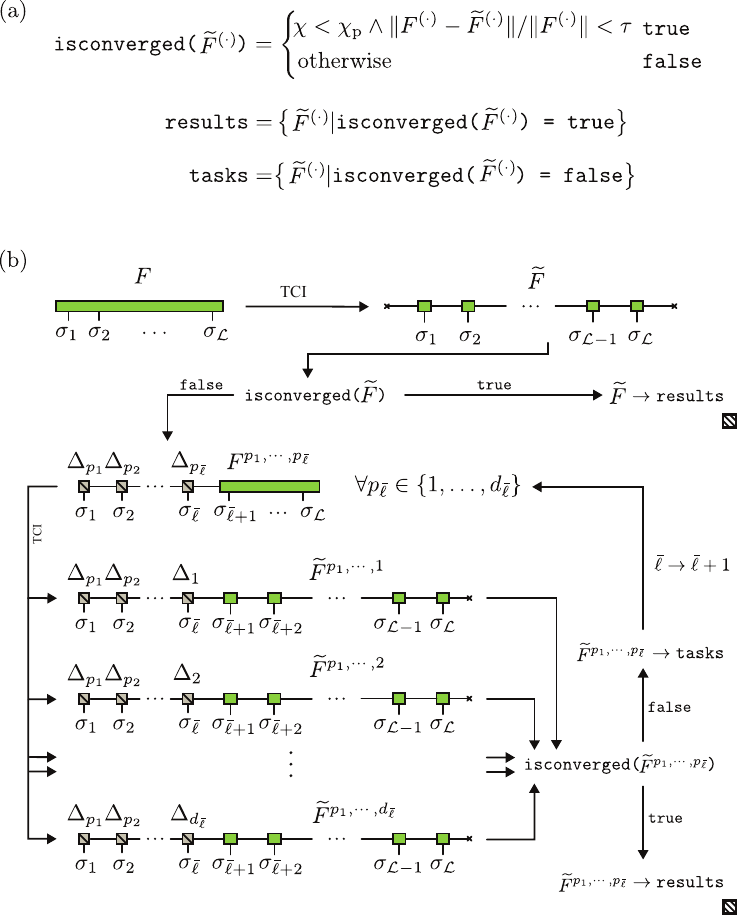}
        \caption{
            Flowchart of the adaptive patching algorithm.
            (a) Convergence criterion for the subtensors $\widetilde{F}^{p_1,\dots,p_{\bar \ell}}$ -- or \textit{patches} -- defined by the parameters $\chip$ and $\tau$, and the two sets of subtensors that have already converged -- \texttt{results} -- and those yet to converge  -- \texttt{tasks}.
            (b) Flowchart of the \textit{patching scheme}. The TT unfolding is adaptively decomposed into smaller computations through \textit{slicing}. 
            Each subtensor is TCI unfolded within the smaller domain.
            $d_{\bar \ell}$ is the number of possible values for the index $p_{\bar \ell}$ (local dimension of the ${\bar \ell}$-th index).
            The converged \textit{patches} are added to \texttt{results}, the yet-to-converge ones to \texttt{tasks}. The algorithm terminates -- \includegraphics[scale=0.95]{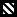} -- when \texttt{tasks} is empty. We refer the reader to the main text for additional details.
            }
        \label{fig:patchingAlg}
    \end{figure}

    Given a maximum rank per patch of \(\chip\), a tolerance \(\tau\), and some initial pivots, the algorithm proceeds as follows (see also the flowchart in Fig.~\ref{fig:patchingAlg}):
    \begin{enumerate}
        \item Initialize a TT \(\tilde{F}_\bsigma\) approximating \(F_\bsigma\) using the initial pivots.
        \item Optimize \(\tilde{F}_\bsigma\) using the 2-site TCI algorithm described in Ref.~\cite{Fernandez2024}, with tolerance \(\tau\) and maximum bond dimension \(\chip\).
        \item Check whether the estimated error \(\bigl\|F_\bsigma - \tilde{F}_\bsigma\bigr\|_\infty/\|F_\bsigma\|_\infty\) is smaller than the tolerance \(\tau\). Here, \(\|F_\bsigma\|_\infty\) is approximated by the maximum value of \(F_\bsigma\) over all sampled points \(\bsigma\).
        If it is, the algorithm terminates and returns \(\tilde{F}_\bsigma\).
        \item Otherwise, pick the next index \(\sigma_\ell\) in a predetermined order (here, \(\sigma_1, \ldots, \sigma_\mL\)) along which the domain should be partitioned. For each value of \(p_\ell = 1, \ldots, d_\ell\), call the algorithm recursively on the patch \(F^{p_\ell}_\bsigma\), using the pivots of \(\tilde{F}_\bsigma\) as initial guess (the \texttt{tasks}). Collect the return values of each call in \texttt{results}, and return the resulting collection.
    \end{enumerate}
    
    We describe the output of the algorithm in more detail.
    To simplify the description, we start projections from the first index \(\sigma_1\) to the last index \(\sigma_\mL\).
    The output result of the algorithm is a collection of tensor trains of the form 
    \begin{equation}
        \raisebox{-4.5mm}{\includegraphics{Plots/PatchTensorTrain.pdf}},
        \label{eq:patchedTTApp}
    \end{equation}
    where both the \textit{prefix} indices $[p_1,\dots,p_{\bar{\ell}}]$ and the length of the prefix -- the \textit{patching level} -- $\bar{\ell} \in \{1,\dots,\mL\}$ can take very different values. Nevertheless, two conditions are respected in our implementation:
    \begin{itemize}
        \item any subtensor contained in \texttt{tasks}  is discarded after further subdomain projection, i.e. all final pairs of index prefixes of the TT in \texttt{results} satisfy
        \begin{equation}
            [p_1,\dots,p_{\bar{\ell}_1}] \nsubseteq [p_1,\dots,p_{\bar{\ell}_2}] \quad \forall\, \bar{\ell}_1 \neq \bar{\ell}_2
        \end{equation}
         Consequently, the adaptive patching procedure produces patches $\widetilde{F}^{p_1,\dots,p_{\bar \ell}}$ that are strictly \textit{non-overlapping} (\textit{disjoint}) in configuration space; 
        \item the collection of TTs of the form in \eqref{eq:patchedTTApp}, in order to constitute a good approximation for the full tensor $F_\bsigma$, satisfies the condition  
        \begin{equation}
            F_\bsigma \approx \sum_{\widetilde{F}^{(\cdot)} \in \texttt{results}} \widetilde{F}^{(\cdot)}.
        \end{equation}
        and we can therefore resum them to a single tensor train $\widetilde{F}_\bsigma$, similar to what we did in \eqref{eq:NslicesSum}. $\widetilde{F}_\bsigma$ is a good TT representation of $F_\bsigma$. However due to the adaptivity of the algorithm, differently from \eqref{eq:NslicesSum}, here we might be presented with a series of \textit{patches} $\widetilde{F}^{p_1,\dots,p_{\bar \ell}}$ with prefixes $(p_1,\dots,p_{\bar \ell})$ of different lengths ${\bar \ell}$; nonetheless, this intricacy does not affect the TT sum operation.
    \end{itemize}


\section{Example illustrating  overpatching}
\label{app:overpatching}

\begin{figure}[htbp]
	\centering
	\includegraphics{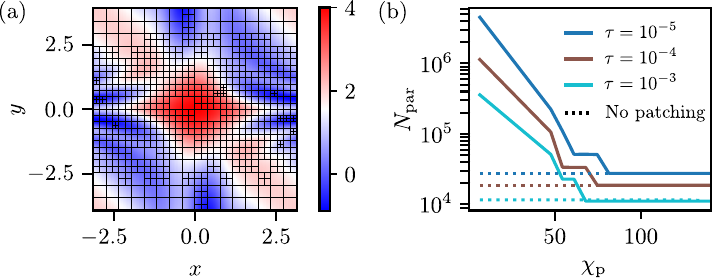}
	\caption{Performance of patched QTCI on the two-dimensional oscillatory function of \eqref{eq:2Dfunction}.  (a) Example of the domain subdivision produced with $\mR = 15$, $\chip = 10$, and $\tau = 10^{-4}$.  The partition is highly redundant and fails to align with the true feature layout of the function. (b) Total number of TT parameters returned by pQTCI (solid curves) at three target tolerances, plotted against the bond-cap $\chip$.  For comparison, the parameter count of a full-domain TCI compression is shown as a dotted line.  Once $\chip\gtrsim 80 = \chi^{\textrm{TCI}}_{\textrm{max}}$, every run collapses to a single patch and matches the TCI cost; for smaller bond-caps, overpatching inflates the parameter count sharply.
    }
	\label{fig:2DOverpatching}
\end{figure}

Fig.~\ref{fig:2DOverpatching} shows how patched QTCI can fail when confronted with a function whose features are \emph{uniformly spread across the whole domain}.  
The target function is 
\begin{align}
\label{eq:2Dfunction}
f(x,y) &=
  1
  + e^{-0.4\,(x^{2}+y^{2})}
  + e^{-x^{2}}\sin(xy)
  + e^{-y^{2}}\cos(3xy) \nonumber \\
  &\quad + \cos(x+y)
  + 0.05\,\cos\bigl[10^{2}\,(2x-4y)\bigr] \\
  &\quad + 5\!\times\!10^{-4}\,\cos\bigl[10^{3}\,(-2x+7y)\bigr]
  + 10^{-5}\,\cos\bigl(2\times10^{8}x\bigr). \nonumber
\end{align}

\section{Prototype algorithm for finding optimal patching order} \label{app:patchordering}

We present a concrete implementation of the heuristic strategy introduced in Sec.~\ref{sec:patch-ordering}, aimed at dynamically identifying a near-optimal patch ordering within the adaptive patching framework. The algorithm is designed for use with the TCI-based adaptive patching routine, pQTCI. Refinements are conceivable, but are left for future work.

\begin{algorithm}[htbp]
  \caption{Identify optimal patching site for a TT.}
  \label{alg:choose-site-with-pivots}
  \DontPrintSemicolon
  \KwIn{
    Tensor $F_{\bsigma}$ (black-box access to elements); \\
    TCI unfolding $\widetilde{F}$ with local bond dimensions $\{\chi_\ell\}_{\ell=1}^{\mL}$ and pivot matrices $\{P_\ell\}_{\ell=1}^{\mL}$; \\
  }

  \KwOut{Chosen splitting site $\ell^\star$.}

  \tcp{1. Find a site with maximal bond dimension}
  $\chi_{\max} \gets \max_{\ell=1,\dots,L} \chi_\ell$\;
  $\ell_{\max} \gets \arg\min\{\ell \mid \chi_\ell = \chi_{\max}\}$ 

  \tcp{2. For each site, reconstruct modified pivot matrices and compute rank cost}
  \For{$\ell \gets 1$ \KwTo $\mL$}{
    $S_\ell \gets 0$\;

    \For{$v \gets 0$ \KwTo $d_\ell - 1$}{
      \tcp{Construct pivot matrix $P(\ell,v)$ with site $\ell$ fixed to $v$}
      Initialise $P(\ell,v) \in \mathbb{C}^{N_{\mathrm{row}} \times N_{\mathrm{col}}}$\;

      \For{$r \gets 1$ \KwTo $\chi_{\max}$}{
        \For{$c \gets 1$ \KwTo $\chi_{\max}$}{
          \tcp{Start from original pivot multi-indices at $\ell_{\max}$}
          $i \gets i_{\ell_{\max}}^{(r)} \in \mathcal{I}_{\ell_{\max}}$\;
          $j \gets j_{\ell_{\max}+1}^{(c)} \in \mathcal{J}_{\ell_{\max}+1}$\;

          \tcp{Reconstruct full index $\boldsymbol{\sigma}$ from $(i,j)$}
          $\boldsymbol{\sigma} \gets i \oplus j $

          \tcp{Overwrite site $\ell$ with fixed value $v$}
          $\widetilde{\bsigma} \gets \bsigma$\;
          $\widetilde{\sigma}_\ell \gets v$\;

          \tcp{Evaluate the full tensor at this modified index}
          $P(\ell,v)_{r,c} \gets F_{\widetilde{\bsigma}}$\;
        }
      }

      \tcp{Compute numerical rank with accuracy $\tau$ of the reconstructed pivot matrix}
      $r_{\ell,v} \gets \mathrm{Rank}_{\tau}\bigl(P(\ell,v)\bigr)$\;

      \tcp{Accumulate squared ranks for site $\ell$}
      $S_\ell \gets S_\ell + r_{\ell,v}^2$\;
    }
  }
  \tcp{3. Choose the site with minimal total squared-rank cost}
  $\ell^\star \gets \arg\min_{\ell=1,\dots,\mL} S_\ell/\mathrm{Rank}_{\tau}^2\bigl(P_{\ell_{\max}}\bigr)$\;

  \Return{$\ell^\star$}\;
\end{algorithm}
The procedure exploits the pivot and pivot-matrix construction of Ref.~\cite{Fernandez2024} to estimate the cost of performing the next patching step at a given TT site $\ell$. The algorithm scans all sites and, for each candidate $ \ell$, constructs a cumulative rank cost by modifying the current pivot matrix at that site. This is achieved by fixing the local index $\sigma_\ell$ of all pivot elements to each admissible value $v = 0, \dots, d_\ell - 1$, yielding a set of modified pivot matrices $P(\ell,v)$.

The numerical rank of each $P(\ell,v)$ is computed and accumulated into a site-dependent cost $S_\ell$, which estimates the rank of the TT that would result from patching at site $\ell$. The optimal splitting site $\ell^\star$ is then chosen by minimizing this cumulative rank cost on all the sites.

By selecting the site with minimal cumulative rank, the algorithm favors patching steps that maximally reduce the rank of the parent TT after patching. When applied at every iteration of the adaptive patching routine, this procedure dynamically constructs a near-optimal patch ordering that adapts to the evolving intermediate tensor trains. Since different TTs at the same patching level may admit different optimal splitting sites, the resulting patch ordering naturally acquires a tree-like structure.

The site-selection step of the patch ordering optimization is detailed below in Algorithm~\ref{alg:choose-site-with-pivots}.

\section{Influence of patch pattern on MPO--MPO contractions}
\label{app:patchPatternMPOMPOContr}
In this appendix, we illustrate in detail the observation made in Sec.~\ref{sec:complexity-elementwise} regarding the dependence of the patched MPO–MPO contraction cost on the patching pattern adopted for the input MPOs. We analyze three representative patching strategies to perform the contraction $\widetilde{C}_{\bsigma\bsigma''} = \sum_{\bsigma'} \widetilde{A}_{\bsigma\bsigma'} \widetilde{B}_{\bsigma'\bsigma''},$ which span the range from optimal to unfavorable configurations.
\begin{itemize}
    \item \textit{Best case.} As discussed in Sec.~\ref{sec:complexity-elementwise}, this scenario corresponds to patching only the internal indices (i.e., the row indices of $\widetilde{A}$ and columns of $\widetilde{B}$). Assuming conservation of the total number of parameters by setting $\chip = \chi / \sqrt{\Np}$, where $\chi = \chi_{\widetilde{A}} = \chi_{\widetilde{B}}$ denotes the maximum bond dimension of the unpatched MPOs approximating $A$ and $B$, the resulting contraction cost scales as $\mathcal{O}(\chi^4\mL/\Np)$
    \item \textit{Worst case.}  In this configuration, only the external MPO indices (corresponding to \(\bsigma\) and \(\bsigma''\)) are patched. Assuming the same number of patches for \(A\) and \(B\), all patch pairs are compatible, yielding \(N_{\textrm{pair}} = \Np^2\). As a result, the contraction cost increases to $\mathcal{O}(\chi^4 \mathcal{L})$ compared with the best case, which matches the scaling of a standard MPO–MPO contraction. In this case, patching provides no computational benefit and introduces only additional overhead.
    \item \textit{General case.} Here, both internal and external indices are patched. Assuming a uniform patching pattern that produces square-like patches in the matrix representation, and an equal number of patches for \(\widetilde{A}\) and \(\widetilde{B}\), the number of compatible patch pairs is $N_{\mathrm{pair}} = d^{\ellb} d^{\ellb (D-1)/D} = \Np^{(2D-1)/D}$, where \(\Np = d^{\ellb}\) and \(\ellb\) denotes the last projected site in both MPOs. For functions of dimensionality \(D=2\) in the interleaved representation ($d=2$), this leads to an overall contraction cost scaling as $ \mathcal{O}(\chi^4 \mathcal{L}/\sqrt{\Np})$, which still offers a computational advantage over the unpatched MPO–MPO contraction.
\end{itemize}

We validate the above complexity estimates through numerical experiments by computing the integral
\begin{equation}
h(x,y) = \int_{-2}^{2} \mathrm{d}s , f(x,s), g(s,y),
\label{eq:testPatchPatterIntegral}
\end{equation}
which corresponds to a continuous matrix multiplication between the two functions $f$ and $g$.
The integrands are defined by 
\begin{align}
    f(x,s) & = a\sum^4_{i=1}e^{-\frac{(x-x_i)^2 +(s-s_i)^2}{\sigma^2_i}} \label{eq:integrandPatchPattern_f}\\ 
    g(s,y) & = a\sum^4_{j=1}e^{-\frac{\sqrt{(s-s_j)^2 +(y-y_j)^2}}{\sigma^2_j}}, \label{eq:integrandPatchPattern_g}
\end{align}
where $a=10^3$, $\sigma_{i/j} = 2^{-(i/j+1)}$, $(x_i, s_i) = (\cos \phi_i, \sin \phi_i)$, $\phi_i = (2i-1)\frac{\pi}{4}$ and $(s_j, y_j) = (x_j +2\sigma_j,s_i + \sigma_j)$. 

\begin{figure}[htbp]
    \centering
    \includegraphics{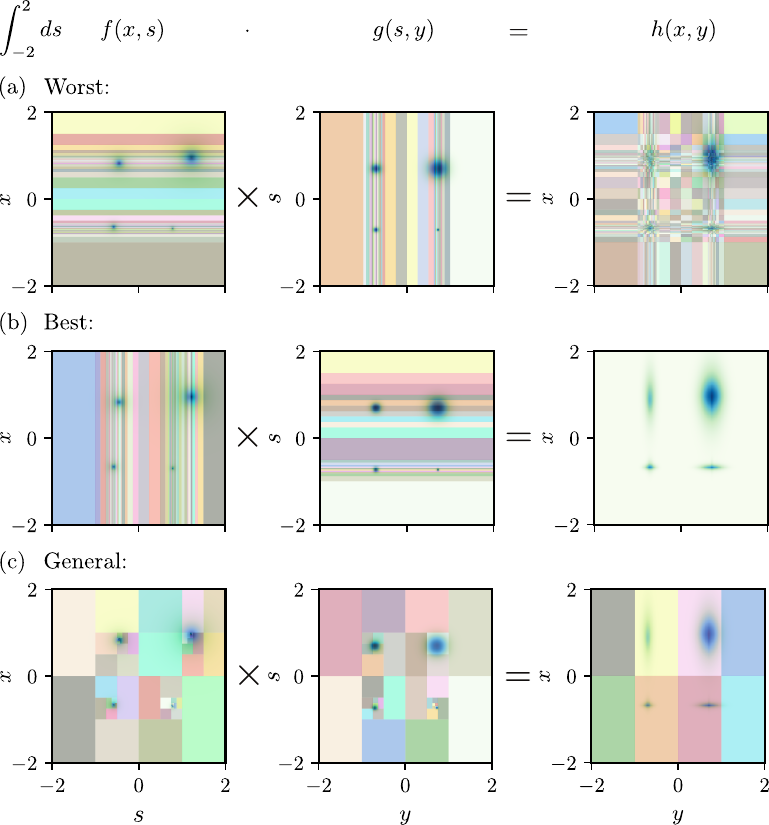}
    \caption{Patch tilings of the two factors \(f(x,s)\) (Eq.~\eqref{eq:integrandPatchPattern_f}) and \(g(s,y)\) (Eq.~\eqref{eq:integrandPatchPattern_g}), together with the resulting product \(h(x,y)\). Each row corresponds to a different patching strategy for the MPO–MPO contraction: (a) row patching against column patching, representing the \emph{worst-case} scenario in terms of the number of contracted patch pairs; (b) column patching against row patching, corresponding to the \emph{best-case}; and (c) interleaved patching, representing the \emph{general} case. The patched result inherits the row and column patch structure of \(f\) and \(g\), respectively. Distinct colors are used to highlight individual patches. }
    \label{fig:patchingPatternsMatMul}
\end{figure}

The integral in Eq.~\eqref{eq:testPatchPatterIntegral} is evaluated using the patched MPO–MPO contraction routines as illustrated in Fig.~\ref{fig:patchingPatternsMatMul}. The functions $f$ and $g$ are patched according to different grid patterns in order to isolate the three contraction scenarios discussed above: the \emph{worst} case (Fig.~\ref{fig:patchingPatternsMatMul}(a)), the \emph{best} case (Fig.~\ref{fig:patchingPatternsMatMul}(b)), and the \emph{general} case (Fig.~\ref{fig:patchingPatternsMatMul}(c)).

\begin{figure}[htbp]
    \centering
    \includegraphics{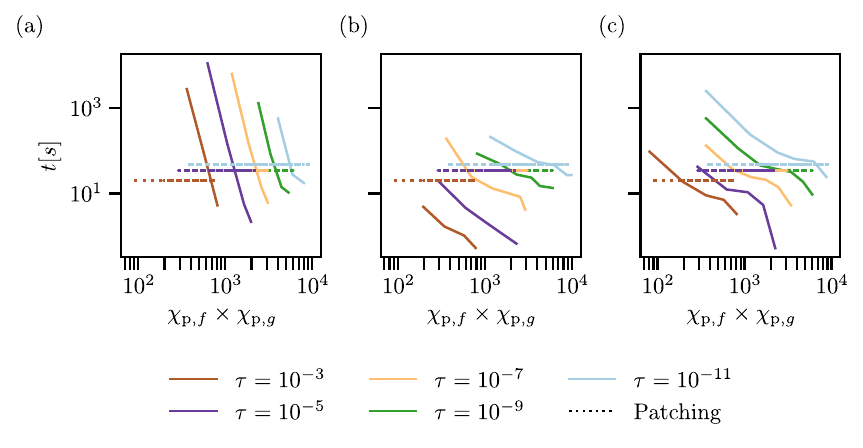}
    \caption{Runtime scaling of the patched MPO--MPO contraction of $f$ and $g$ defined in Eqs.~\eqref{eq:integrandPatchPattern_f}-\eqref{eq:integrandPatchPattern_g}, shown for the (a) worst-case, (b) best-case, and (c) general patching strategies. The functions are represented with QTTs with interleaved ordering of bits and \(\mathcal{R}=17\), before mapping them to MPOs. Results are reported for different choices of the bond-caps $\chi_{\textrm{p},f}$ and $\chi_{\textrm{p},g}$. Dotted lines indicate the reference runtime of a monolithic MPO--MPO contraction performed at the same tolerance. }
    \label{fig:patchedMulResults}
\end{figure}

Figure~\ref{fig:patchedMulResults} presents a benchmark comparison of the three patching strategies discussed above. The functions are approximated using a QTT representation with interleaved ordering and $\mR=17$, before mapping them to MPOs. While varying the bond-caps, $\chi_{\textrm{p},f}$ and $\chi_{\textrm{p},g}$, used in the patched approximations of $f$ and $g$, we measure the runtime of the corresponding patched MPO–MPO contractions on an Intel Xeon W-2245 CPU @ 3.90\,GHz. The three panels correspond to the \emph{worst} (Fig.~\ref{fig:patchedMulResults}(a)), \emph{best} (Fig.~\ref{fig:patchedMulResults}(b)), and \emph{general} (Fig.~\ref{fig:patchedMulResults}(c)) patching configurations for the input MPOs.
In all cases, the measured runtimes are compared against the cost of a standard (monolithic) MPO–MPO contraction of the same integrands at identical tolerance. As anticipated, the optimal patching strategy shown in Fig.~\ref{fig:patchedMulResults}(b) consistently yields the lowest runtime, outperforming the other configurations. In the worst-case scenario (Fig.~\ref{fig:patchedMulResults}(a)), some parameter regimes still exhibit some speedup over the monolithic contraction; this effect is likely attributable to the reduced memory footprint reached by adaptive patching when the integrands are spatially localized (no constraint on preserving the total number of parameters in the patched approximation for numerical results). The general configuration in Fig.~\ref{fig:patchedMulResults}(c) exhibits intermediate performance, in agreement with the theoretical scaling predictions.

Despite its limited scope, the benchmark agrees with our theoretical predictions and underscores the importance of the patching pattern in determining contraction efficiency, opening the door to future work on optimizing contractions through tailored patching strategies.
\clearpage

\end{appendix}

\bibliography{main.bib}

@article{BickersFLEX,
  title = {Fluctuation-exchange theory for general lattice Hamiltonians},
  author = {Esirgen, G\"okhan and Bickers, N. E.},
  journal = {Phys. Rev. B},
  volume = {55},
  issue = {4},
  pages = {2122--2143},
  numpages = {0},
  year = {1997},
  month = {Jan},
  publisher = {American Physical Society},
  doi = {10.1103/PhysRevB.55.2122},
  url = {https://link.aps.org/doi/10.1103/PhysRevB.55.2122}
}

@article{Aryasetiawan1998,
doi = {10.1088/0034-4885/61/3/002},
url = {https://doi.org/10.1088/0034-4885/61/3/002},
year = {1998},
month = {mar},
publisher = {},
volume = {61},
number = {3},
pages = {237},
author = {F Aryasetiawan and O Gunnarsson},
title = {The  GW method},
journal = {Reports on Progress in Physics}
}

@article{Vilk1997,
	author = {{Y.M. Vilk} and {A.-M.S. Tremblay}},
	title = {Non-Perturbative Many-Body Approach to the Hubbard Model and Single-Particle Pseudogap},
	DOI= "10.1051/jp1:1997135",
	url= "https://doi.org/10.1051/jp1:1997135",
	journal = {J. Phys. I France},
	year = 1997,
	volume = 7,
	number = 11,
	pages = "1309-1368",
	month = "",
}

@article{Fannes1992,
author={Fannes, M.
and Nachtergaele, B.
and Werner, R. F.},
title={Finitely correlated states on quantum spin chains},
journal={Communications in Mathematical Physics},
year={1992},
month={Mar},
day={01},
volume={144},
number={3},
pages={443-490},
issn={1432-0916},
doi={10.1007/BF02099178},
url={https://doi.org/10.1007/BF02099178}
}

@article{White1992,
  title = {Density matrix formulation for quantum renormalization groups},
  author = {White, Steven R.},
  journal = {Phys. Rev. Lett.},
  volume = {69},
  issue = {19},
  pages = {2863--2866},
  numpages = {0},
  year = {1992},
  month = {11},
  publisher = {American Physical Society},
  doi = {10.1103/PhysRevLett.69.2863},
  url = {https://link.aps.org/doi/10.1103/PhysRevLett.69.2863}
}

@article{Schollwock2011,
title = {The density-matrix renormalization group in the age of matrix product states},
journal = {Annals of Physics},
volume = {326},
number = {1},
pages = {96-192},
year = {2011},
note = {January 2011 Special Issue},
issn = {0003-4916},
doi = {https://doi.org/10.1016/j.aop.2010.09.012},
url = {https://www.sciencedirect.com/science/article/pii/S0003491610001752},
author = {Ulrich Schollwöck}
}

@article{Vidal2003,
  title = {Efficient Classical Simulation of Slightly Entangled Quantum Computations},
  author = {Vidal, Guifr\'e},
  journal = {Phys. Rev. Lett.},
  volume = {91},
  issue = {14},
  pages = {147902},
  numpages = {4},
  year = {2003},
  month = {10},
  publisher = {American Physical Society},
  doi = {10.1103/PhysRevLett.91.147902},
  url = {https://link.aps.org/doi/10.1103/PhysRevLett.91.147902}
}

@article{VerstraeteCirac2004,
      title={Renormalization algorithms for Quantum-Many Body Systems in two and higher dimensions}, 
      author={F. Verstraete and J. I. Cirac},
      archivePrefix={arXiv},
      year={2004},
      primaryClass={cond-mat.str-el},
      url={https://arxiv.org/abs/cond-mat/0407066}, 
      note={\href{https://arxiv.org/abs/cond-mat/0407066}{arXiv:0407066 [cond-mat]}}
}

@article{Lindsey2024,
      title={Multiscale interpolative construction of quantized tensor trains}, 
      author={Michael Lindsey},
      year={2023},
      archivePrefix={arXiv},
      primaryClass={math.NA},
      url={https://arxiv.org/abs/2311.12554}, 
      note={\href{https://arxiv.org/abs/2311.12554}{arXiv:2311.12554 [math.Na]}}
}

@misc{vonDelftTNNotes,
    author       = {von Delft, Jan},
    title        = {{Lecture Notes on Tensor Networks for Many-Body Physics}},
    note = {Ludwig-Maximilians University of Munich, Theoretical Solid State Physics, \href{https://www2.physik.uni-muenchen.de/lehre/vorlesungen/sose_23/tensor_networks_23/skript/index.html}{Lecture Notes (2023)}},
}

@misc{tensor4all.org,
  note        = {Tensor4all, \href{https://tensor4all.org}{tensor4all.org}},
}

@article{Fernandez2024,
 author        = {N\'u\~nez Fern\'andez, Yuriel and Marc K. Ritter and Matthieu Jeannin and Jheng-Wei Li and Thomas Kloss and Thibaud Louvet and Satoshi Terasaki and Olivier Parcollet and Jan von Delft and Hiroshi Shinaoka and Xavier Waintal},
 date          = {2025},
 title         = {Learning tensor networks with tensor cross interpolation: new algorithms and libraries},
 journal       = {SciPost Phys.},
 volume        = {18},
 pages         = {104},
 doi           = {10.21468/SciPostPhys.18.3.104},
 url           = {https://scipost.org/SciPostPhys.18.3.104},
}

@article{Fernandez2022,
  title = {{Learning Feynman Diagrams with Tensor Trains}},
  author = {N\'u\~nez Fern\'andez, Yuriel and Jeannin, Matthieu and Dumitrescu, Philipp T. and Kloss, Thomas and Kaye, Jason and Parcollet, Olivier and Waintal, Xavier},
  journal = {Phys. Rev. X},
  volume = {12},
  issue = {4},
  pages = {041018},
  numpages = {30},
  year = {2022},
  month = {11},
  publisher = {American Physical Society},
  doi = {10.1103/PhysRevX.12.041018},
  url = {https://link.aps.org/doi/10.1103/PhysRevX.12.041018}
}

@article{Michailidis2025TTM,
  title = {Element-Wise Multiplication of Tensor Trains},
  author = {Michailidis, Alexios A. and Fenton, Christian and Kiffner, Martin},
  journal = {SIAM Journal on Scientific Computing},
  volume = {47},
  number = {5},
  pages = {B1158--B1174},
  year = {2025},
  doi = {10.1137/24M1714149},
}

@article{Oseledets2010,
title = {{TT-cross approximation for multidimensional arrays}},
journal = {Linear Algebra and its Applications},
volume = {432},
number = {1},
pages = {70-88},
year = {2010},
issn = {0024-3795},
doi = {https://doi.org/10.1016/j.laa.2009.07.024},
url = {https://www.sciencedirect.com/science/article/pii/S0024379509003747},
author = {Ivan Oseledets and Eugene Tyrtyshnikov}
}

@article{Oseledets2011,
author = {Oseledets, I. V.},
title = {Tensor-Train Decomposition},
journal = {SIAM Journal on Scientific Computing},
volume = {33},
number = {5},
pages = {2295-2317},
year = {2011},
doi = {10.1137/090752286},

URL = { 
    
        https://doi.org/10.1137/090752286
    
    

}
}

@article{Dolgov2020,
title = {Parallel cross interpolation for high-precision calculation of high-dimensional integrals},
journal = {Computer Physics Communications},
volume = {246},
pages = {106869},
year = {2020},
issn = {0010-4655},
doi = {https://doi.org/10.1016/j.cpc.2019.106869},
url = {https://www.sciencedirect.com/science/article/pii/S0010465519302565},
author = {Sergey Dolgov and Dmitry Savostyanov}
}

@article{Ritter2024,
  title = {Quantics Tensor Cross Interpolation for High-Resolution Parsimonious Representations of Multivariate Functions},
  author = {Ritter, Marc K. and N\'u\~nez Fern\'andez, Yuriel and Wallerberger, Markus and von Delft, Jan and Shinaoka, Hiroshi and Waintal, Xavier},
  journal = {Phys. Rev. Lett.},
  volume = {132},
  issue = {5},
  pages = {056501},
  numpages = {6},
  year = {2024},
  month = {Jan},
  publisher = {American Physical Society},
  doi = {10.1103/PhysRevLett.132.056501},
  url = {https://link.aps.org/doi/10.1103/PhysRevLett.132.056501}
}

@article{Jolly2024,
  title = {Tensorized orbitals for computational chemistry},
  author = {Jolly, Nicolas and N\'u\~nez Fern\'andez, Yuriel and Waintal, Xavier},
  journal = {Phys. Rev. B},
  volume = {111},
  issue = {24},
  pages = {245115},
  numpages = {12},
  year = {2025},
  month = {Jun},
  publisher = {American Physical Society},
  doi = {10.1103/PhysRevB.111.245115},
  url = {https://link.aps.org/doi/10.1103/PhysRevB.111.245115}
}

@article{Oseledets2009,
	author = {Oseledets, I. V.},
	doi = {10.1134/S1064562409050056},
	id = {Oseledets2009},
	isbn = {1531-8362},
	journal = {Doklady Mathematics},
	number = {2},
	pages = {653--654},
	title = {Approximation of matrices with logarithmic number of parameters},
	url = {https://doi.org/10.1134/S1064562409050056},
	volume = {80},
	year = {2009}}

@article{Khoromskij2011,
	author = {Khoromskij, Boris N. },
	doi = {10.1007/s00365-011-9131-1},
	id = {Khoromskij2011},
	isbn = {1432-0940},
	journal = {Constructive Approximation},
	number = {2},
	pages = {257--280},
	title = {O(dlog N)-Quantics Approximation of N-d Tensors in High-Dimensional Numerical Modeling},
	url = {https://doi.org/10.1007/s00365-011-9131-1},
	volume = {34},
	year = {2011}}

@article{Hiroshi2023,
  title = {Multiscale Space-Time Ansatz for Correlation Functions of Quantum Systems Based on Quantics Tensor Trains},
  author = {Shinaoka, Hiroshi and Wallerberger, Markus and Murakami, Yuta and Nogaki, Kosuke and Sakurai, Rihito and Werner, Philipp and Kauch, Anna},
  journal = {Phys. Rev. X},
  volume = {13},
  issue = {2},
  pages = {021015},
  numpages = {27},
  year = {2023},
  month = {Apr},
  publisher = {American Physical Society},
  doi = {10.1103/PhysRevX.13.021015},
  url = {https://link.aps.org/doi/10.1103/PhysRevX.13.021015}
}

@article{Murray2024,
  title = {Nonequilibrium diagrammatic many-body simulations with quantics tensor trains},
  author = {Murray, Matthias and Shinaoka, Hiroshi and Werner, Philipp},
  journal = {Phys. Rev. B},
  volume = {109},
  issue = {16},
  pages = {165135},
  numpages = {12},
  year = {2024},
  month = {Apr},
  publisher = {American Physical Society},
  doi = {10.1103/PhysRevB.109.165135},
  url = {https://link.aps.org/doi/10.1103/PhysRevB.109.165135}
}

@article{Oseledets2009Intro,
author = {Oseledets, I. V. and Tyrtyshnikov, E. E.},
title = {Breaking the Curse of Dimensionality, Or How to Use SVD in Many Dimensions},
journal = {SIAM Journal on Scientific Computing},
volume = {31},
number = {5},
pages = {3744-3759},
year = {2009},
doi = {10.1137/090748330},

URL = { 
    
        https://doi.org/10.1137/090748330
    
    

}
}

@article{Kolda2009,
author = {Kolda, Tamara G. and Bader, Brett W.},
title = {Tensor Decompositions and Applications},
journal = {SIAM Review},
volume = {51},
number = {3},
pages = {455-500},
year = {2009},
doi = {10.1137/07070111X},

URL = { 
    
        https://doi.org/10.1137/07070111X
    
    

}
}

@article{ITensors.jl,
	title={{The ITensor Software Library for Tensor Network Calculations}},
	author={Matthew Fishman and Steven R. White and E. Miles Stoudenmire},
	journal={SciPost Phys. Codebases},
	volume={4},
	year={2022},
	publisher={SciPost},
	doi={10.21468/SciPostPhysCodeb.4},
	url={https://scipost.org/10.21468/SciPostPhysCodeb.4},
}

@article{QSpace,
	title={{QSpace - An open-source tensor library for Abelian and non-Abelian symmetries}},
	author={Andreas Weichselbaum},
	journal={SciPost Phys. Codebases},
	volume={40},
	year={2024},
	publisher={SciPost},
	doi={10.21468/SciPostPhysCodeb.40},
	url={https://scipost.org/10.21468/SciPostPhysCodeb.40},
}

@misc{TensorCrossInterpolation.jl,
	author  = {Ritter Marc and Shinaoka Hiroshi and Satoshi Terasaki},
	title   = {TensorCrossInterpolation.jl},
	url     = {https://github.com/tensor4all/TensorCrossInterpolation.jl/},
	version = {v0.9.0},
  note = {\href{https://github.com/tensor4all/TensorCrossInterpolation.jl/}{https://github.com/tensor4all/TCI.jl}}

}

@article{Stoudenmire2010,
doi = {10.1088/1367-2630/12/5/055026},
url = {https://dx.doi.org/10.1088/1367-2630/12/5/055026},
year = {2010},
month = {may},
publisher = {},
volume = {12},
number = {5},
pages = {055026},
author = {Stoudenmire, E M and White, Steven R},
title = {Minimally entangled typical thermal state algorithms},
journal = {New Journal of Physics}
}

@article{Thunstrom2018,
  title = {Analytical investigation of singularities in two-particle irreducible vertex functions of the Hubbard atom},
  author = {Thunstr\"om, P. and Gunnarsson, O. and Ciuchi, Sergio and Rohringer, G.},
  journal = {Phys. Rev. B},
  volume = {98},
  issue = {23},
  pages = {235107},
  numpages = {16},
  year = {2018},
  month = {Dec},
  publisher = {American Physical Society},
  doi = {10.1103/PhysRevB.98.235107},
  url = {https://link.aps.org/doi/10.1103/PhysRevB.98.235107}
}

@article{RevModPhys.90.025003,
  title = {Diagrammatic routes to nonlocal correlations beyond dynamical mean field theory},
  author = {Rohringer, G. and Hafermann, H. and Toschi, A. and Katanin, A. A. and Antipov, A. E. and Katsnelson, M. I. and Lichtenstein, A. I. and Rubtsov, A. N. and Held, K.},
  journal = {Rev. Mod. Phys.},
  volume = {90},
  issue = {2},
  pages = {025003},
  numpages = {53},
  year = {2018},
  month = {May},
  publisher = {American Physical Society},
  doi = {10.1103/RevModPhys.90.025003},
  url = {https://link.aps.org/doi/10.1103/RevModPhys.90.025003}
}

@misc{Rohshap2025b,
      title={Diagnosing phase transitions through time scale entanglement}, 
      author={Stefan Rohshap and Hirone Ishida and Frederic Bippus and Anna Kauch and Karsten Held and Hiroshi Shinaoka and Markus Wallerberger},
      year={2025},
      archivePrefix={arXiv},
      primaryClass={cond-mat.str-el},
      url={https://arxiv.org/abs/2507.11276}, 
      note={\href{https://arxiv.org/abs/2507.11276}{arXiv:2507.11276 [cond-mat.str-el]}}
}

@article{Rohshap2025a,
  title = {Two-particle calculations with quantics tensor trains: Solving the parquet equations},
  author = {Rohshap, Stefan and Ritter, Marc K. and Shinaoka, Hiroshi and von Delft, Jan and Wallerberger, Markus and Kauch, Anna},
  journal = {Phys. Rev. Res.},
  volume = {7},
  issue = {2},
  pages = {023087},
  numpages = {21},
  year = {2025},
  month = {Apr},
  publisher = {American Physical Society},
  doi = {10.1103/PhysRevResearch.7.023087},
  url = {https://link.aps.org/doi/10.1103/PhysRevResearch.7.023087}
}

@article{Rohshap2025c,
  title = {Entanglement across scales: {Q}uantics tensor trains as a natural framework for renormalization},
  author = {Rohshap, Stefan and Li, Jheng-Wei and Lorenz, Alena and Hasil, Serap and Held, Karsten and Kauch, Anna and Wallerberger, Markus},
  journal = {Phys. Rev. Res.},
  volume = {7},
  issue = {4},
  pages = {043313},
  numpages = {31},
  year = {2025},
  month = {Dec},
  publisher = {American Physical Society},
  doi = {10.1103/qlrs-6f8t},
  url = {https://link.aps.org/doi/10.1103/qlrs-6f8t}
}

@Article{Dolgov2012,
  author    = {Dolgov, Sergey and Khoromskij, Boris and Savostyanov, Dmitry},
  journal   = {Journal of Fourier Analysis and Applications},
  title     = {Superfast {F}ourier Transform Using {QTT} Approximation},
  year      = {2012},
  issn      = {1531-5851},
  month     = may,
  number    = {5},
  pages     = {915--953},
  volume    = {18},
  doi       = {10.1007/s00041-012-9227-4},
  publisher = {Springer Science and Business Media LLC},
}

@ARTICLE{Gourianov2022-fg,
  title   = "A Quantum Inspired Approach to Exploit Turbulence Structures",
  author  = "Gourianov, Nikita and Lubasch, Michael and Dolgov, Sergey and van
             den Berg, Quincy Y and Babaee, Hessam and Givi, Peyman and Kiffner,
             Martin and Jaksch, Dieter",
  journal = "Nature Computational Science",
  volume  =  2,
  pages   = "30--37",
  year    =  2022,
  url     = "https://www.nature.com/articles/s43588-021-00181-1",
  doi     = "10.1038/s43588-021-00181-1"
}

@misc{tensorkitjl,
      title={TensorKit.jl: A Julia package for large-scale tensor computations, with a hint of category theory}, 
      author={Lukas Devos and Jutho Haegeman},
      archivePrefix={arXiv},
      year={2025},
      primaryClass={cs.MS},
      url={https://arxiv.org/abs/2508.10076}, 
      note={\href{https://arxiv.org/abs/2508.10076}{arXiv:2508.10076 [cs.MS]}}
}

@article{Sroda2024-jm,
  title = {Memory-Efficient Nonequilibrium Green's Function Framework Built On Quantics Tensor Trains},
  author = {\ifmmode \acute{S}\else \'{S}\fi{}roda, Maksymilian and Inayoshi, Ken and Shinaoka, Hiroshi and Werner, Philipp},
  journal = {Phys. Rev. Lett.},
  volume = {135},
  issue = {22},
  pages = {226501},
  numpages = {8},
  year = {2025},
  month = {Nov},
  publisher = {American Physical Society},
  doi = {10.1103/dxfb-b3l5},
  url = {https://link.aps.org/doi/10.1103/dxfb-b3l5}
}

@BOOK{Barth2005-jf,
  title     = "Adaptive Mesh Refinement - Theory and Applications",
  author    = "Barth, Timothy J and Griebel, Michael and Keyes, David E and
               York, New and Nieminen, Risto M and Roose, Dirk and Leuven, Tamar
               and Schlick, New",
  publisher = "Springer Berlin, Heidelberg",
  year      =  2005,
  isbn      =  9783540211471
}

@Book{Qttbook,
  Title                    = {Tensor Numerical Methods in Scientific Computing},
  Author                   = {B. N. Khoromskij},
  Publisher                = {De Gruyter, Berlin, Boston},
  Year                     = {2018},
  Edition                  = {First},
  Series                   = {Radon Series on Computational and Applied Mathematics},
  Volume                   = {19},

  Doi                      = {10.1515/9783110365917}
}

@ARTICLE{XTRGthermal,
  title   = "Exponential Thermal Tensor Network Approach for Quantum Lattice
             Models",
  author  = "{Bin-Bin Chen and Lei Chen and Ziyu Chen and Wei Li and
             Andreas Weichselbaum}",
  journal = "PHYSICAL REVIEW X",
  volume  =  8,
  pages   =  031082,
  year    =  2018,
  url     = "https://link.aps.org/doi/10.1103/PhysRevX.8.031082",
  doi     = "10.1103/PhysRevX.8.031082"
}

@ARTICLE{Andrade2021-ws,
  title     = "{GRChombo}: An adaptable numerical relativity code for
               fundamental physics",
  author    = "Andrade, Tomas and Salo, Llibert and Aurrekoetxea, Josu and
               Bamber, Jamie and Clough, Katy and Croft, Robin and de Jong, Eloy
               and Drew, Amelia and Duran, Alejandro and Ferreira, Pedro and
               Figueras, Pau and Finkel, Hal and Frana, Tiago and Ge, Bo-Xuan
               and Gu, Chenxia and Helfer, Thomas and J{\"{a}}ykk{\"{a}}, Juha
               and Joana, Cristian and Kunesch, Markus and Kornet, Kacper and
               Lim, Eugene and Muia, Francesco and Nazari, Zainab and Radia,
               Miren and Ripley, Justin and Shellard, Paul and Sperhake, Ulrich
               and Traykova, Dina and Tunyasuvunakool, Saran and Wang, Zipeng
               and Widdicombe, James and Wong, Kaze",
  journal   = "J. Open Source Softw.",
  publisher = "The Open Journal",
  volume    =  6,
  number    =  68,
  pages     =  3703,
  month     =  dec,
  year      =  2021,
  url       = "http://dx.doi.org/10.21105/joss.03703",
  doi       = "10.21105/joss.03703",
  issn      = "2475-9066"
}

@ARTICLE{Zhang2021-qp,
  title     = "{AMReX}: Block-structured adaptive mesh refinement for
               multiphysics applications",
  author    = "Zhang, Weiqun and Myers, Andrew and Gott, Kevin and Almgren, Ann
               and Bell, John",
  journal   = "Int. J. High Perform. Comput. Appl.",
  publisher = "SAGE Publications",
  volume    =  35,
  number    =  6,
  pages     = "508--526",
  month     =  nov,
  year      =  2021,
  url       = "http://dx.doi.org/10.1177/10943420211022811",
  doi       = "10.1177/10943420211022811",
  issn      = "1094-3420,1741-2846",
  language  = "en"
}

@ARTICLE{Drew2023-oh,
  title     = "Radiation from global topological strings using adaptive mesh
               refinement: Massive modes",
  author    = "Drew, Amelia and Shellard, E P S",
  journal   = "Phys. Rev. D.",
  publisher = "American Physical Society (APS)",
  volume    =  107,
  number    =  4,
  pages     =  043507,
  month     =  feb,
  year      =  2023,
  url       = "http://dx.doi.org/10.1103/PhysRevD.107.043507",
  doi       = "10.1103/physrevd.107.043507",
  issn      = "2470-0010,2470-0029",
  language  = "en"
}

@Article{Tyrtyshnikov1996,
author={Tyrtyshnikov, Eugene},
title={Mosaic-Skeleton approximations},
journal={CALCOLO},
year={1996},
month={Jun},
day={01},
volume={33},
number={1},
pages={47-57},
issn={1126-5434},
doi={10.1007/BF02575706},
url={https://doi.org/10.1007/BF02575706}
}

@Article{Gavrilyuk2002,
  author    = {Gavrilyuk, Ivan P. and Hackbusch, Wolfgang and Khoromskij, Boris N.},
  title     = {\(\mathcal{H}\)-Matrix approximation for the operator exponential with applications},
  journal   = {Numerische Mathematik},
  year      = {2002},
  volume    = {92},
  number    = {1},
  pages     = {83--111},
  month     = jul,
  doi       = {10.1007/s002110100360},
  issn      = {0945-3245},
  url       = {https://doi.org/10.1007/s002110100360}
}

@ARTICLE{Drew2022-qg,
  title     = "Radiation from global topological strings using adaptive mesh
               refinement: Methodology and massless modes",
  author    = "Drew, Amelia and Shellard, E P S",
  journal   = "Phys. Rev. D.",
  publisher = "American Physical Society (APS)",
  volume    =  105,
  number    =  6,
  pages     =  063517,
  month     =  mar,
  year      =  2022,
  url       = "http://dx.doi.org/10.1103/PhysRevD.105.063517",
  doi       = "10.1103/physrevd.105.063517",
  issn      = "2470-0010,2470-0029",
  language  = "en"
}

@misc{inayoshi2025causalitybaseddivideandconqueralgorithmnonequilibrium,
      title={A causality-based divide-and-conquer algorithm for nonequilibrium Green's function calculations with quantics tensor trains}, 
      author={Ken Inayoshi and Maksymilian Środa and Anna Kauch and Philipp Werner and Hiroshi Shinaoka},
      year={2025},
      archivePrefix={arXiv},
      primaryClass={cond-mat.str-el},
      url={https://arxiv.org/abs/2509.15028}, 
      note={\href{https://arxiv.org/abs/2509.15028}{arXiv:2509.15028 [cond-mat.str-el]}}
}

@ARTICLE{Kaye2021-ly,
  title     = "Low rank compression in the numerical solution of the
               nonequilibrium Dyson equation",
  author    = "Kaye, Jason and Golez, Denis",
  journal   = "SciPost Phys.",
  publisher = "Stichting SciPost",
  volume    =  10,
  number    =  4,
  pages     =  091,
  month     =  apr,
  year      =  2021,
  url       = "https://scipost.org/10.21468/SciPostPhys.10.4.091",
  doi       = "10.21468/scipostphys.10.4.091",
  issn      = "2542-4653"
}

@misc{chen2025solvinggrosspitaevskiiequationquantic,
      title={Solving the Gross-Pitaevskii Equation with Quantic Tensor Trains: Ground States and Nonlinear Dynamics}, 
      author={Qian-Can Chen and I-Kang Liu and Jheng-Wei Li and Chia-Min Chung},
      year={2025},
      archivePrefix={arXiv},
      primaryClass={cond-mat.quant-gas},
      url={https://arxiv.org/abs/2507.04279}, 
      note = {\href{https://arxiv.org/abs/2507.04279}{arXiv:2507.04279 [cond-mat.quant-gas]}}
}

@misc{boucomas2025quanticstensortrainsolving,
      title={Quantics Tensor Train for solving Gross-Pitaevskii equation}, 
      author={Aleix Bou-Comas and Marcin Płodzień and Luca Tagliacozzo and Juan José García-Ripoll},
      year={2025},
      archivePrefix={arXiv},
      primaryClass={cond-mat.quant-gas},
      url={https://arxiv.org/abs/2507.03134}, 
      note = {\href{https://arxiv.org/abs/2507.03134}{arXiv:2507.03134 [cond-mat.quant-gas]}}
}

@article{sun2025selfconsistenttensornetworkmethod,
  title = {Self-consistent tensor network method for correlated super-moir\'e matter beyond one billion sites},
  author = {Sun, Yitao and Niedermeier, Marcel and Ant\~ao, Tiago V. C. and Fumega, Adolfo O. and Lado, Jose L.},
  journal = {Phys. Rev. Res.},
  volume = {7},
  issue = {4},
  pages = {043288},
  numpages = {14},
  year = {2025},
  month = {Dec},
  publisher = {American Physical Society},
  doi = {10.1103/krjp-mn4v},
  url = {https://link.aps.org/doi/10.1103/krjp-mn4v}
}

@misc{meng_recursive_2026,
      title={Recursive Sketched Interpolation: Efficient Hadamard Products of Tensor Trains}, 
      author={Zhaonan Meng and Yuehaw Khoo and Jiajia Li and E. Miles Stoudenmire},
      year={2026},
      eprint={2602.17974},
      archivePrefix={arXiv},
      primaryClass={quant-ph},
      url={https://arxiv.org/abs/2602.17974},
      note={\href{https://arxiv.org/abs/2602.17974}{arXiv:2602.17974 [quant-ph]}}
}

@article{ritter_elementwise_2026,
    author = {Marc K.\ Ritter},
    title = {Elementwise operations on tensor trains},
    journal = {to be published},
    year = {2026},
}
\end{document}